\documentclass[preprintnumbers,amsmath,11pt,amssymb,floatfix,superscriptaddress,nofootinbib]{article}
\usepackage{relsize}
\usepackage{cite}
\usepackage{graphicx}
\begin{document}
\Large{\textbf{THERMODYNAMICS OF BLACK HOLES IN RAINBOW GRAVITY}}\\
\begin{center}
Ritwick Banerjee$^{*1}$ and Ritabrata Biswas$^{\dagger 2}$ \\
\vspace{1cm}
$*$Department of Mathematics, Narasinha Dutt College, Howrah, India\\
$\dagger$ Department of Mathematics, Bankura University, Bankura, India 
\end{center}
\begin{center}
\vspace{2cm}
\Large{\textbf{Abstract}}
\end{center}
In this paper, we investigate the thermodynamic properties of black holes under the influence of
rainbow gravity. In the metric of Schwarzschild, Reissner-Nordstrom and Reissner-Nordstrom-
de Sitter black hole surrounded by quintessence, we consider a rainbow function and derive
the existence of remnant and critical masses of a black hole. Using the Hawking temperature
relation we derive the heat capacity and the entropy of the rainbow gravity inspired black holes
and closely study the relation between entropy and area of the horizon for different values of n
of the rainbow function. \\
\vspace{1cm}

Keywords $:$ Black hole physics, Rainbow gravity, Thermodynamics.\\
\vspace{1cm}
\section{Introduction}

The discussion about the Lorentz symmetry at Planck scale leads us to many possible answers. Keeping the central physical message of theory of relativity unchanged, namely the equivalence of all inertial observers, we can propose double/deformed special relativity(DSR). Two postulates of relativity in that case can be formulated as : the equivalence of all inertial observers are taken and secondly, assumption of two observer independent scales : one speed of light $c$ and the other is the dimension of mass $k$ (or length $\lambda=k^{-1}$), identified with the Planck mass. In the limit $k\rightarrow \infty$, DSR becomes special relativity \cite{1,2,3,4}. Now in such a quantum phenomenological area near the Planck scale, the standard energy momentum dispersion relations are modified. Magueijo and Smolin \cite{5,6} have extended the DSR to general relativity. Their proposition was the energy of a test particle with the background geometry and consequently the modified dispersion relation as :\\
\begin{equation}
E^{2}f\left( \frac{E}{E_p} \right)^{2}-p^2 g\left( \frac{E}{E_p} \right)^{2}=m^2
\end{equation}
here $p$, $m$ and $E_p$ are the momentum, the mass of the test particle and the Planck energy. So different background geometry will be observed by differently energised quantas. This is why we call it Rainbow gravity. Literature is enriched by works related to gravity at the Planck scale \cite{7,8,9,10,11,12,13,14,15,16,17,18,19,20,21,22,23,24,25,26}. The nature of rainbow functions have been discussed in many existing literatures. \cite{27,28,29,30,31,32,33,34,35}.
Popular forms are : \\
\begin{equation} \label{Rainbow}
f\left( \frac{E}{E_p} \right)=1 ~~~ , ~~~~ g\left( \frac{E}{E_p} \right)=\sqrt{1-\eta\left(\frac{E}{E_p}\right)^n}
\end{equation}
here $n$ is a positive integer and $\eta$ is a constant of order unity. Both the functions become unity while $E\rightarrow 0$.\\
The Rainbow gravity inspired BHs give us a deeper insight of the fate of BH evaporation. When the heat capacity vanishes we can say that the BH evaporation stops and it gives us the remnant mass of the BH. Also it can be observed that the thermodynamic outcomes for the Einstein and Rainbow gravity BHs are more or less similar. From this we can say that the laws of physics are equivalent for both the cases \cite{35}. Recently, Gim \& Kim \cite{36} have shown that Schwarzschild BH in Rainbow gravity in an isothermal cavity 
additional Hawking page phase transition near the event horizon apart from the standard one giving rise to the idea of existence of local BH.  \\

It can be observed that the modification of metric by certain popular forms of rainbow functions, changes the thermodynamical behaviour of the different black holes. The modification changes the temperature and the entropy of the systems and it brings forward the ideas of critical mass and the remnant mass of the black holes from the thermodynamic point of view. Thus the rainbow function in a way prevents the complete evaporation of a black hole leaving behind a remnant mass which is exactly the same way as done by the generalized uncertainty principle (GUP). For the chosen rainbow function the entropy of the system have a lot of similarity to those derived using the GUP. \\

The main motivations for studying black holes under rainbow gravity are as follows. Due to the high energy levels of black holes, it is important to study the properties of black holes after considering the quantum corrections on the classical perspectives. The idea of energy dependent spacetime is one of those quantum corrections. Considering this energy dependent spacetime, we can venture on the effects it brings about on the thermodynamic properties of the black hole. One of the noticeable effect is the existence of remnant mass of a black hole which can be proposed to be a major candidate for solving the information paradox \cite{37} and also being UV completion of Einstein gravity \cite{6}. Also, we come across a critical mass which shows a second order phase transition making the stable black hole unstable. To have a better picture about the thermodynamic properties of the black holes it is essential to consider the high energy regimes. Here we consider the rainbow gravity to study its effects on the thermodynamic properties of the black holes.\\

In this paper we have organised as follows. In section $2$ we first talk about the basics of rainbow gravity and the way it modifies the metric of a Schwarzschild black hole and the other thermodynamic constraints. Then we go on to find the critical mass, the remnant mass and finally the entropy of the system under the influence of rainbow gravity. Meanwhile in the process of deriving these meaningful constraints we try to gain deeper insights into behaviour of the different thermodynamic constraints by plotting the temperature vs. mass, heat capacity vs. mass and entropy vs. area of horizon of rainbow gravity inspired Schwarzschild black hole. We plot these graphs for different values of $\eta$ and $n$ which gives us a beter scope of comparison. Then we move on to section $3$ and $4$ where we discuss the same results as before for Reissner Nordstrom and Reissner Nordstrom de Sitter black hole surrounded by quintessence. Finaly, we conclude in section $5$.\\
\section{Thermodynamics of Rainbow Gravity inspired Schwarzschild black hole}
In this section we want to study the different thermodynamic properties of Schwarzschild BH taking into account the effect of rainbow gravity functions.  The Schwarzschild black hole metric inspired by rainbow gravity is given by (\ref{Rainbow}) 
\begin{equation}\label{Rainbow_Sch_metric}
ds^2= -\frac{1}{f^2(\frac{E}{E_{p}})}\left(1-\frac{2MG}{r}\right)dt^2 + \frac{1}{g^2(\frac{E}{E_{p}})}\left(1-\frac{2MG}{r}\right)^{-1}dr^2 + \frac{r^2}{g^2(\frac{E}{E_{p}})}d\Omega^2 \\
\end{equation}
We relate surface gravity $\kappa$ to the Hawking temperature by the relation $T=\frac{\kappa}{2\pi}$ and  the surface gravity is defined by $\kappa=\lim_{r \to R_{s}} \sqrt{-\frac{1}{4}g^{rr}g^{tt}(g_{tt,r})^2}$ ~~ where $R_{s}=2GM$ is the Schwarzschild radius.\\ From (\ref{Rainbow_Sch_metric}) we get \\
$g^{tt}=-f^2(\frac{E}{E_{p}})\left(1-\frac{2MG}{r}\right)^{-1},$ $\ \ \ g^{rr}=g^2(\frac{E}{E_{p}})\left(1-\frac{2MG}{r}\right),$ $\ \ \ (g_{tt,r})^2=\frac{1}{f^4(\frac{E}{E_{p}})}\left(\frac{2MG}{r^2}\right)^2$ \\
Hence, the surface gravity of the Schwarzschild black hole under the effect of rainbow gravity is given by 
\begin{equation}\label{Surface_gravity_rainbow}
\kappa=\frac{g(\frac{E}{E_{p}})}{f(\frac{E}{E_{p}})}\frac{1}{4MG}
\end{equation}
Therefore, the Hawking temperature is given by 
\begin{equation}\label{Rainbow_gravity_Temperature}
T=\frac{1}{8\pi G}\sqrt{\frac{1}{M^2}-\frac{\eta}{(2GE_{p})^n}\frac{1}{M^{n+2}}}
\end{equation}
In the above expression we have set $E=\frac{1}{2GM}$. Equation (\ref{Rainbow_gravity_Temperature}) gives us a relation between the Temperature and Mass of rainbow gravity inspired Schwarzschild BH.
\begin{figure}[h]
\begin{center}
Fig.$1a$ \ ~~~\hspace{6.5cm} Fig.$1b$~~~~~\\
\includegraphics[height=2.5in, width=3.2in]{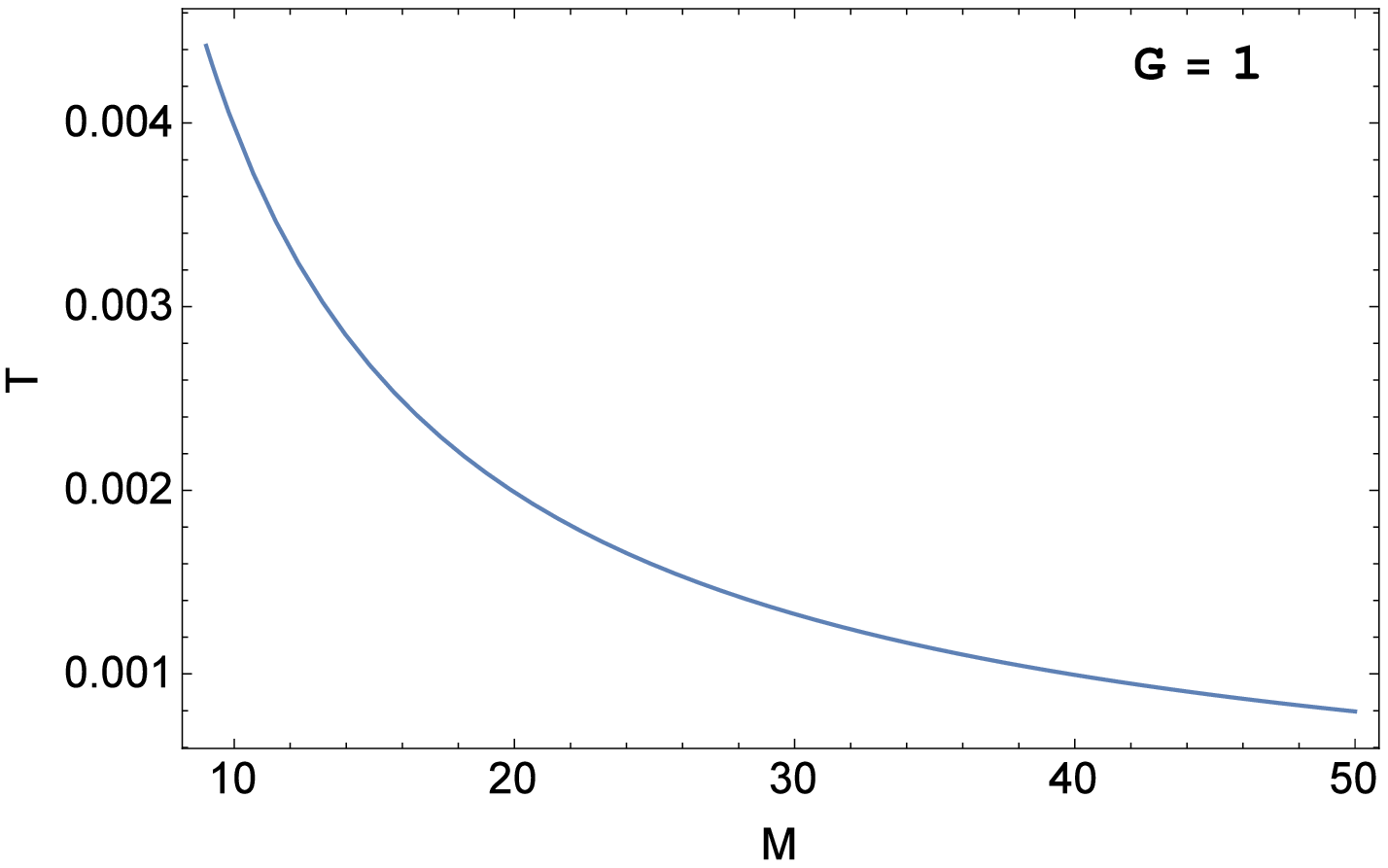}~~~~
\vspace{.1cm}
\includegraphics[height=2.5in, width=3.2in]{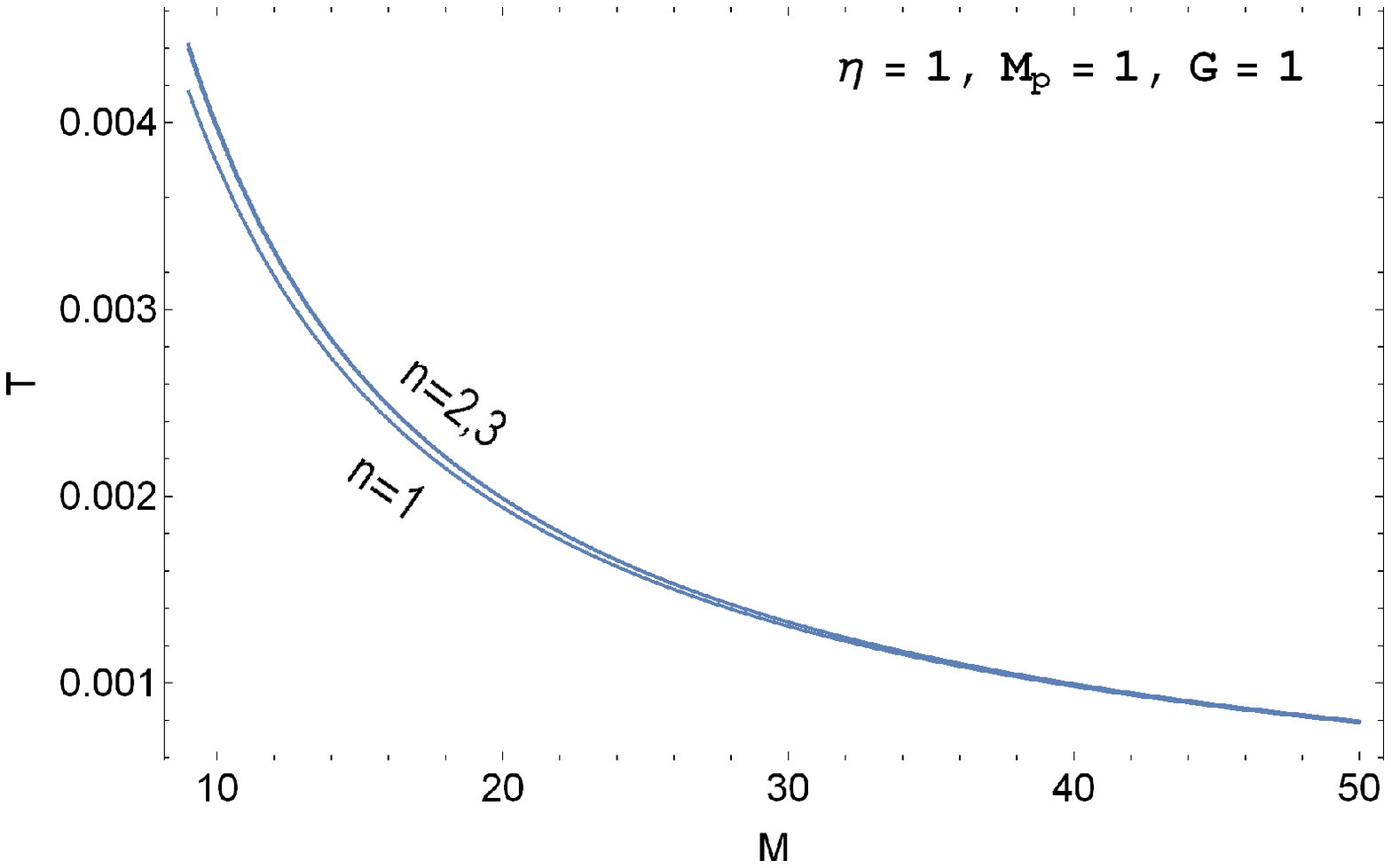}~~\\
Fig.$1a$ and $1b$ represents the Temperature vs Mass of the black hole curves for Schwarzschild black holes in Einstein and Rainbow gravity respectively.
\end{center}
\end{figure}\\
Figure $1a$ represents the curve of Hawking temperature of Schwarzschild black hole in Einstein gravity vs the mass of the same. It shows that for low mass the temperature is high and as we increase the mass, temperature graph reduces and becomes asymptotic to the $M$ axis. Now introduction to rainbow gravity keeps the general trend almost same. It has been represented by figure $1b$. But one thing to be noted is that if we count $n$ to be one then for low mass, temperature is comparatively lower than the case of $n=2,3$.
Now $\frac{dM}{T}=dS$. So if for the same mass we have lower $T$ the entropy is greater, that indicates $n=1$ represents higher entropy system. So $n=1$ carries higher disturbances than $n=2,3$ case. \\
For the temperature to be a real quantity we must have \\
\begin{equation}
\frac{1}{M^2}-\frac{\eta}{(2GE_{p})^n}\frac{1}{M^{n+2}}\geq 0 .
\end{equation}
The above condition gives rise to a critical mass $(M_{cr})$. Below the critical mass, the
temperature is not a real quantity. This critical mass is given by :
\begin{equation}
M_{cr}=\frac{\eta^{\frac{1}{n}}}{2GE_{p}}=\eta^{\frac{1}{n}}M_{p} .
\end{equation}
where we have taken $E_{p}=\frac{1}{2GM_{p}} \ $ .
From (\ref{Rainbow_gravity_Temperature}) we get \\
\begin{center}
$ \frac{dT}{dM}=\frac{\left(\frac{(n+2)\eta}{(2GE_{p})^n}\frac{1}{M^{n+3}}-\frac{2}{M^3}\right)}{16\pi G\sqrt{\frac{1}{M^2}-\frac{\eta}{(2GE_{p})^n}\frac{1}{M^{n+2}}}} .
$
\end{center}
The heat capacity of the rainbow gravity inspired Schwarzschild BH is give by
\begin{equation}
C=\frac{dM}{dT}=\frac{16\pi G\sqrt{\frac{1}{M^2}-\frac{\eta}{(2GE_{p})^n}\frac{1}{M^{n+2}}}}{\left(\frac{(n+2)\eta}{(2GE_{p})^n}\frac{1}{M^{n+3}}-\frac{2}{M^3}\right)}.\label{C}
\end{equation}\\
\begin{figure}[h]
\begin{center}
~~~~~~~~~~~~~~~~~~~~~~Fig.2\\
\includegraphics[height=2.5in, width=3.2in]{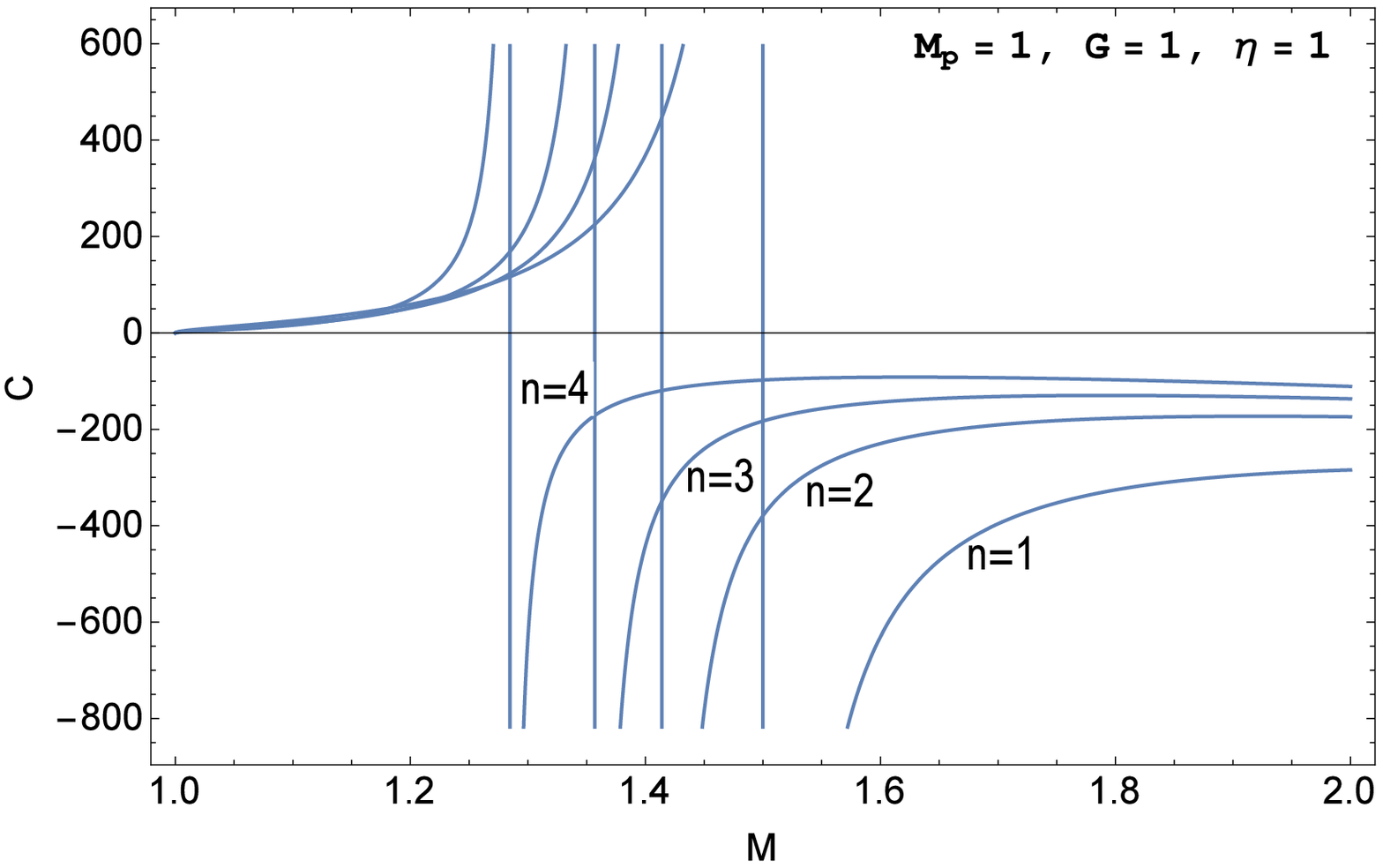}\\
Fig.$2$ represents C vs M curves for $n=1,2,3,4$. For $\eta<1$ the graphs shift to the left side and for $\eta>1$ the graphs shift to the right side keeping their basic tendency same.
\end{center}
\end{figure}\\
It is to be followed that where we do expect to have the critical mass, C vanishes there. The denominator vanishes at a point : \\
$M_{cr_{2}}= \left(\frac{n+2}{2}\right)^{\frac{1}{n}}\frac{\eta^{\frac{1}{n}}}{2G E_{p}}$\\ \\
$=\left( 1+\frac{n}{2} \right)^{\frac{1}{n}}\eta^{\frac{1}{n}}M_p > \eta^{\frac{1}{n}}M_p = M_{cr}$\\ \\
At $M_{cr_{2}}$ we have a second order phase transition which makes the stable black hole unstable. Figure 2 depicts this incident. If we increase $n$, $M_{cr_{2}}$ decreases. i.e., the phase transitions occur faster.
If we increase or decrease $\eta$ the curve completely shifts on the right or left hand side respectively. But for high n such shifting is less.\\
If we set $C=0$, we obtain the remnant mass $M_{rem}$(where the black hole stops evaporating). This gives
\begin{equation}
M_{rem}=\frac{\eta^{\frac{1}{n}}}{2GE_{p}}=\eta^{\frac{1}{n}}M_{p} .
\end{equation}
Thus we primarily can think that the remnant mass of the black hole is equal to its critical
mass. But actually $M_p \eta^{\frac{1}{n}}$ is such a point where the black hole starts its journey. Before that no physical black hole is present. This is why at $M_p \eta^{\frac{1}{n}}$ we get $C=0$. 
Now we can observe that if we take the rainbow gravity parameter $\eta=1$ we get, \\
$M_{rem}=M_{p}$, where $M_p$ is the Planck mass.
From the mass-temperature graph we saw that the temperature was increasing as the mass was decreasing, but when the mass of a BH reaches the remnant mass i.e., the Planck mass, the heat capacity vanishes and the
temperature suddenly becomes zero. So, we can say that at Planck scale the BH evaporation stops and prevents the BH from total evaporation. Hence, the rainbow gravity may solve the information loss and naked singularity problems of black holes.
Now increase of $\eta$ causes increase of $M_{cr}$ as well as $M_{cr_{2}}$. But if n is high $\eta^{\frac{1}{n}}$ tends to one showing no big effect of $\eta 's$ increment.\\
The entropy can be calculated by using the heat capacity of this black hole as follows :\\
\begin{equation}
S=\int C\frac{dT}{T}=\int \frac{dM}{dT}. \label{S}
\end{equation}
Substituting equation (\ref{C}) in equation (\ref{S}) and carrying out a binomial expansion keeping terms upto $O(\eta^{4})$ and assuming that $n\geq 3$ leads to :\\
$ S=8\pi G \mathlarger{\int} \frac{dM}{\sqrt{\frac{1}{M^2}-\frac{\eta}{(2GE_{p})^n}\frac{1}{M^{n+2}}}}\\ \\ $
$ =8\pi G \mathlarger{\int} \ \left[ M + \frac{\eta}{2(2GE_{p})^n}\frac{1}{M^{n-1}} + \frac{3\eta^{2}}{8(2GE_{p})^{2n}}\frac{1}{M^{2n-1}} + \frac{5\eta^{3}}{16(2GE_{p})^{3n}}\frac{1}{M^{3n-1}}+\frac{35\eta^{4}}{128(2GE_{p})^{4n}}\frac{1}{M^{4n-1}}\right]dM $ \\ \\ 
$ =8\pi G \left[ \frac{M^2}{2} + \frac{\eta M^{2-n}}{2(2GE_{p})^n(2-n)} + \frac{3\eta^{2}M^{2-2n}}{8(2GE_{p})^{2n}(2-2n)} + \frac{5\eta^3M^{2-3n}}{16(2GE_{p})^{3n}(2-3n)} + \frac{35\eta^4M^{2-4n}}{128(2GE_{p})^{4n}(2-4n)} \right] \\ \\  $
$
=S_{BH}(GM^2_{p})+\frac{S^{1-\frac{n}{2}}_{BH}\pi^{\frac{n}{2}}\eta (GM^{2-n}_{p})}{(2-n)(GE_{p})^n}+ \frac{3 S^{1-n}_{BH}\pi^n\eta^{2} (GM^{2-2n}_{p})}{8(1-n)(GE_{p})^2n} + \frac{5 S^{1-\frac{3n}{2}}_{BH}\pi^{\frac{3n}{2}}\eta^{3} (GM^{2-3n}_{p})}{8(2-3n)(GE_{p})^3n} +\frac{35S^{1-2n}_{BH}\pi^{2n}\eta^{4} (GM^{2-4n}_{p})}{128(1-2n)(GE_{p})^4n} $\\
\begin{equation}
\resizebox{.97 \textwidth}{!} 
{$=S_{BH}(GM^2_{p})+\frac{S^{1-\frac{n}{2}}_{BH}\pi^{\frac{n}{2}}\eta (2^nGM^{2}_{p})}{(2-n)}+ \frac{3 S^{1-n}_{BH}\pi^n\eta^{2} (2^{2n}GM^{2}_{p})}{8(1-n)} + \frac{5 S^{1-\frac{3n}{2}}_{BH}\pi^{\frac{3n}{2}}\eta^{3} (2^{3n}GM^{2}_{p})}{8(2-3n)}+\frac{35S^{1-2n}_{BH}\pi^{2n}\eta^{4} (2^{4n}GM^{2}_{p})}{128(1-2n)}$
}
\end{equation}
where $S_{BH}=\frac{4\pi M^2}{M^2_{p}}$ is the semi-classical Bekenstein-Hawking entropy for the Schwarzschild black hole. The reason for assuming $n \geq 3$ is that the result of the integration is not valid for $n=1,2$ which can be easily seen from the integrand.\\
In terms of the area of the horizon $A=4\pi R^2_{s} = 16\pi G^2 M^2 = 4l^2_{p}S_{BH}$, the above expression for the entropy can be put in the form :\\
$
S= \left(\frac{A}{4}\right)(GM^2_{p})+\frac{\left(\frac{A}{4}\right)^{1-\frac{n}{2}}\pi^{\frac{n}{2}}\eta (2^nGM^{2}_{p})}{(2-n)}+ \frac{3 \left(\frac{A}{4}\right)^{1-n}\pi^n\eta^{2} (2^{2n}GM^{2}_{p})}{8(1-n)} + \frac{5 \left(\frac{A}{4}\right)^{1-\frac{3n}{2}}\pi^{\frac{3n}{2}}\eta^{3} (2^{3n}GM^{2}_{p})}{8(2-3n)}  \\ $
\begin{equation}
+\frac{35\left(\frac{A}{4}\right)^{1-2n}\pi^{2n}\eta^{4} (2^{4n}GM^{2}_{p})}{128(1-2n)}  \\
\end{equation}\\
where we have set $l_{p}=1.$ \\
We plot $S$ for $\eta=1$ and different values of $n\geq 3$ vs $A$ in Fig.3 and Fig.3.1. We may consider $S\geq 0$ part only to be the physical black hole solutions. It shows as we increase $n$, black hole can even exist for lower area of event horizon. For lower $n$-s $S$ increases directly with $A$. Higher $n$ breaks the curve into two parts - lower $A$ steeply increasing $S$ and after certain point increasing but with a low slope. This shows when black hole is small a change in $A$ causes rapid change in $S$. But later this rapidness decreases.\\ \\ \\ \\
\begin{figure}[h]
\begin{center}
Fig.3~~~~~~~~~~~\hspace{5cm}~~~~~~~~~~~~~~~~~~Fig.3.1\\
\includegraphics[height=2.5in, width=3.2in]{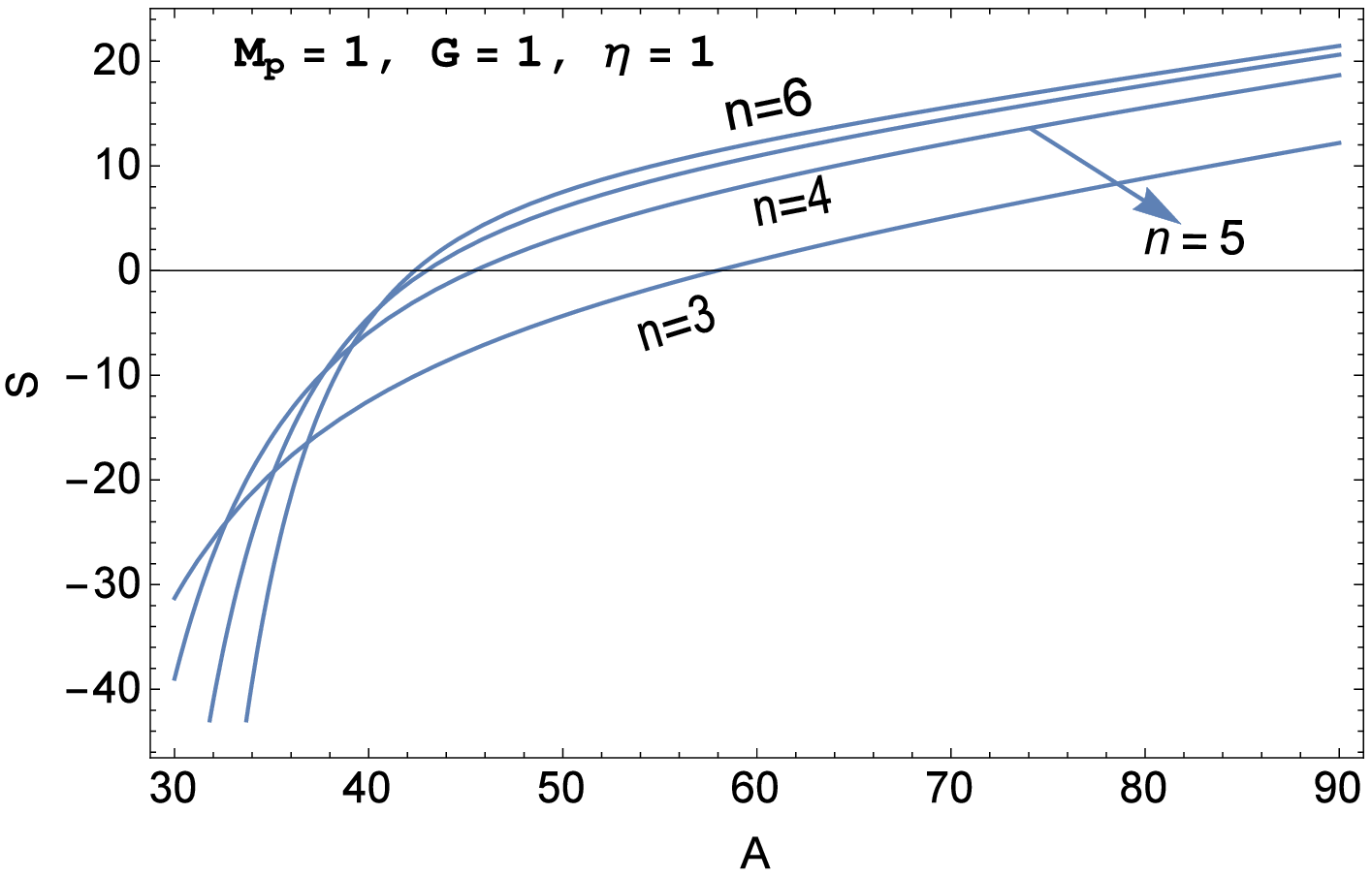} \hspace{1.5cm}
\includegraphics[height=2.5in, width=3.2in]{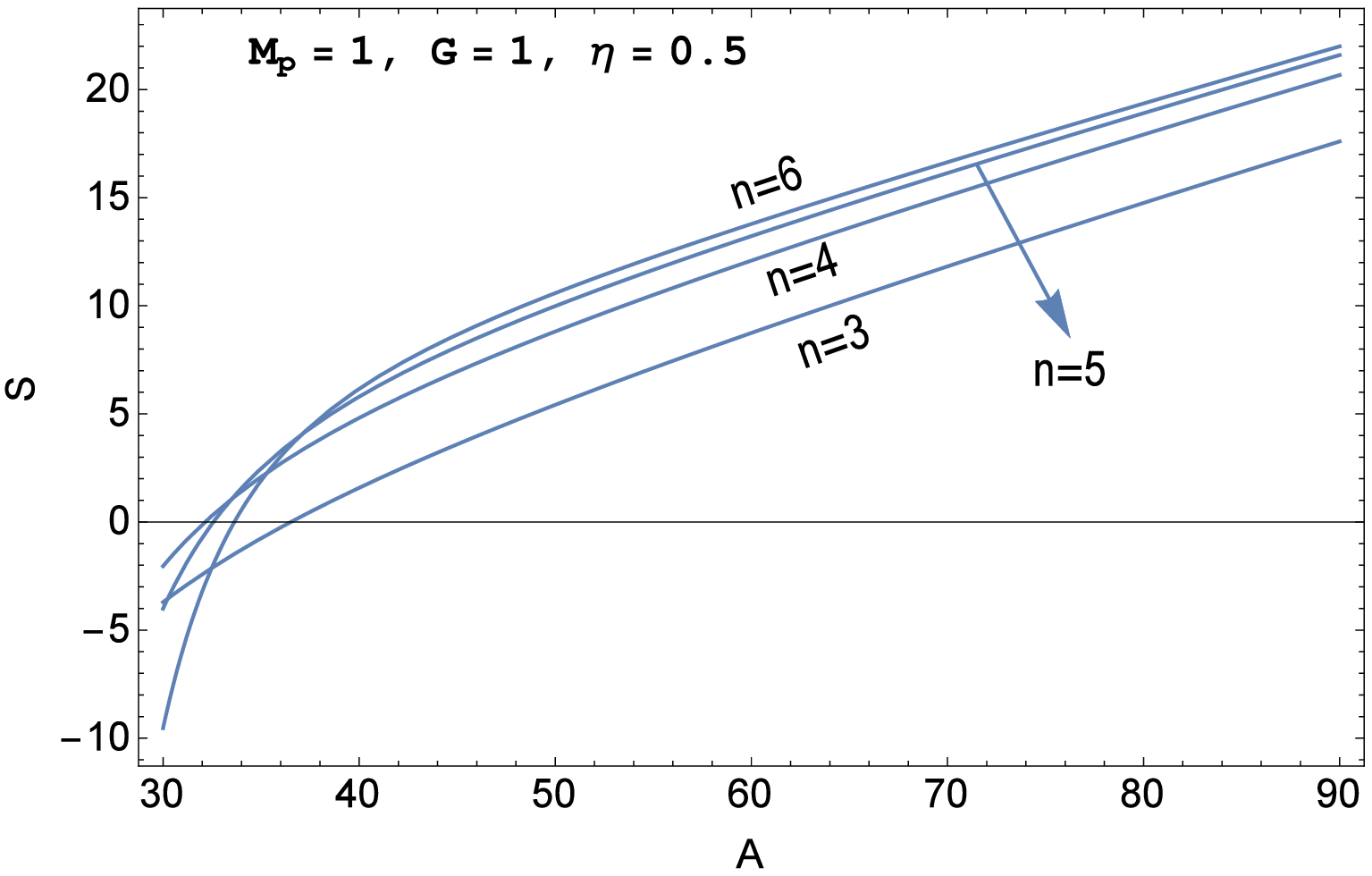}
\end{center}
\end{figure}\\
For $n=1$, the entropy expression upto $O(\eta^{4})$ in terms of horizon area takes the form :\\
$ S=8\pi G \mathlarger{\int} \left[ M + \frac{\eta}{2(2GE_{p})} + \frac{3\eta^{2}}{8(2GE_{p})^{2}}\frac{1}{M} + \frac{5\eta^{3}}{16(2GE_{p})^{3}}\frac{1}{M^{2}}  \\ \\
+\frac{35\eta^{4}}{128(2GE_{p})^{4}}\frac{1}{M^{3}}\right]dM$ \\ \\
$ =8\pi G \left[\frac{M^2}{2}+ \frac{\eta M}{2(2GE_{p})}+ \frac{3\eta^2 lnM}{8(2GE_{p})^2}-\frac{5\eta^3}{16(2GE_{p})^3}\frac{1}{M}- \frac{35\eta^4}{256(2GE_{p})^4}\frac{1}{M^2} \right]$ \\ \\
\resizebox{.97 \textwidth}{!} 
{
$ =S_{BH}(GM^2_{p})+ 2\eta \sqrt{\pi S_{BH}}(GM^2_{p})+\frac{3\pi \eta^2}{2}\left(lnS_{BH}+ln\frac{M^2_{p}}{4\pi}\right)(GM^2_{p})-5S^{\frac{-1}{2}}_{BH}\eta^3\pi^{\frac{3}{2}}(GM^{2}_{p})- \frac{35}{8} S^{-1}_{BH}\eta^4\pi^{2}(GM^2_{p})$
}\\ \\
$=\left(\frac{A}{4}\right)(GM^2_{p})+ 2\eta \sqrt{\pi \left(\frac{A}{4}\right)}(GM^2_{p})+\frac{3\pi \eta^2}{2}\left(ln\left(\frac{A}{4}\right)+ln\frac{M^2_{p}}{4\pi}\right)(GM^2_{p})-5\left(\frac{A}{4}\right)^{\frac{-1}{2}}\eta^3\pi^{\frac{3}{2}}(GM^{2}_{p}) $ \\
\begin{equation}
-\frac{35}{8} \left(\frac{A}{4}\right)^{-1}\eta^4\pi^{2}(GM^2_{p})
\end{equation}\\
For $n=2$, the entropy expression upto $O(\eta^{4})$ in terms of horizon area takes the form :\\ \\
$ S=8\pi G \mathlarger{ \int} \left[ M + \frac{\eta}{2(2GE_{p})^2}\frac{1}{M} + \frac{3\eta^{2}}{8(2GE_{p})^{4}}\frac{1}{M^3} + \frac{5\eta^{3}}{16(2GE_{p})^{6}}\frac{1}{M^{5}}  \\ \\
+\frac{35\eta^{4}}{128(2GE_{p})^{8}}\frac{1}{M^{7}}\right]dM$ \\ \\
$
=8\pi G  \left[\frac{M^2}{2}+ \frac{\eta}{2}M^2_{p}lnM -\frac{3\eta^2M^{4}_{p}}{16M^2}- \frac{5\eta^3M^{6}_{p}}{64M^4}- \frac{35\eta^4M^{8}_{p}}{768M^6} \right]
$\\ \\
\resizebox{.97 \textwidth}{!} 
{
$ = S_{BH}(GM^2_{p})+ 4\eta \left[lnS_{BH} + ln\frac{M^2_{p}}{4\pi }\right](GM^2_{p})-6S^{-1}_{BH}\eta^2 \pi^2 (GM^2_{p}) - 10S^{-2}_{BH}\eta^3 \pi^3 (GM^2_{p})- \frac{70}{3} S^{-3}_{BH}\eta^4 \pi^4 (GM^2_{p})
$
}\\ 
\begin{equation}
\resizebox{.97 \textwidth}{!} 
{
$= \left(\frac{A}{4}\right)(GM^2_{p})+ 4\eta \pi \left[ln\left(\frac{A}{4}\right) + ln\frac{M^2_{p}}{4\pi }\right](GM^2_{p})-6\left(\frac{A}{4}\right)^{-1}\eta^2 \pi^2 (GM^2_{p}) - 10\left(\frac{A}{4}\right)^{-2}\eta^3 \pi^3 (GM^2_{p})-\frac{70}{3} \left(\frac{A}{4}\right)^{-3}\eta^4 \pi^4 (GM^2_{p})$
}
\end{equation}
\begin{figure}[h]
\begin{center}
Fig.3.2~~~~~~~~~~~~~~~\hspace{5cm}~~~~~~~~~~~~~~~~Fig.3.3\\
\includegraphics[height=2.5in, width=3.2in]{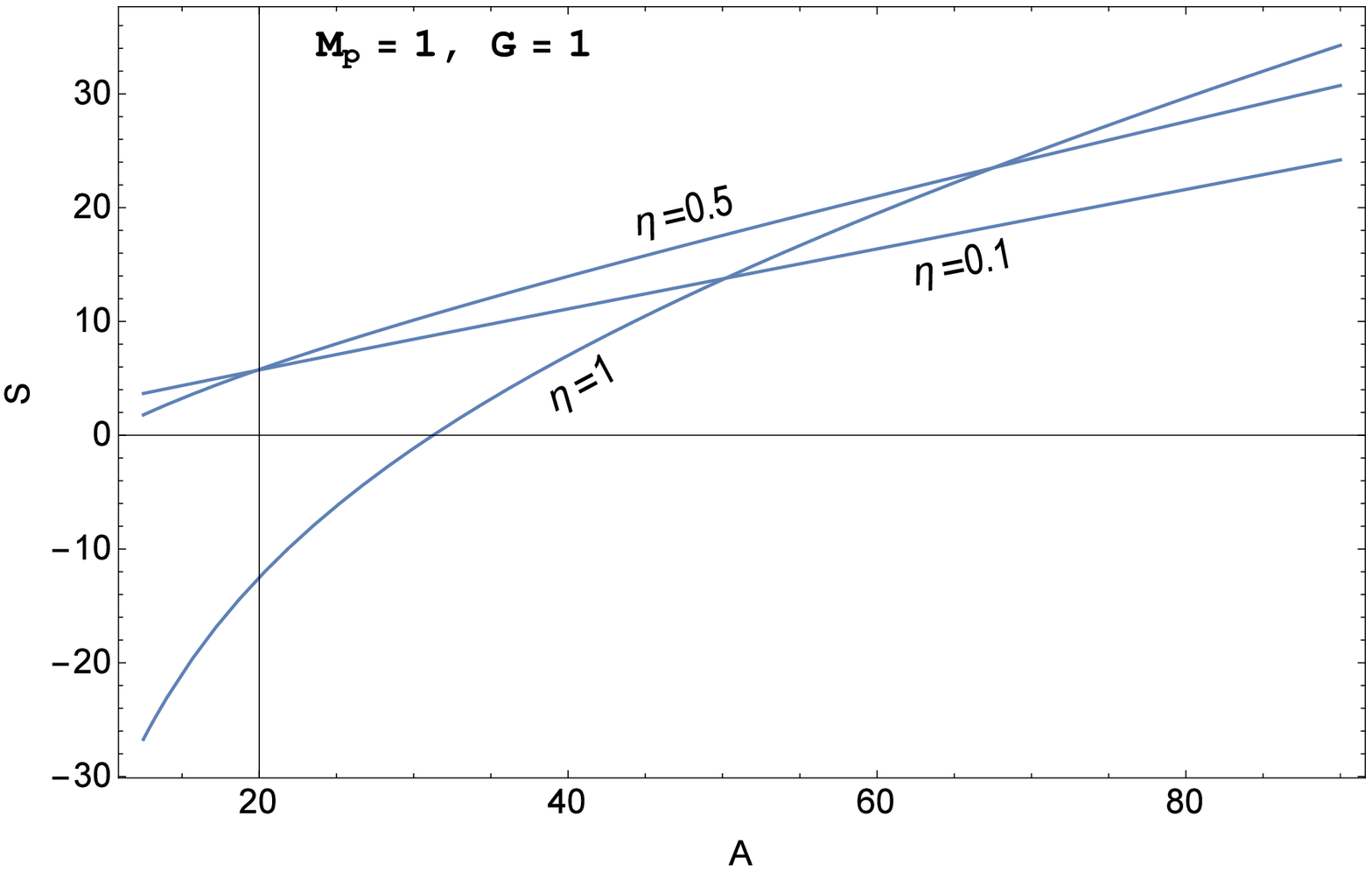} ~~~\hspace{1.5cm}
\includegraphics[height=2.5in, width=3.2in]{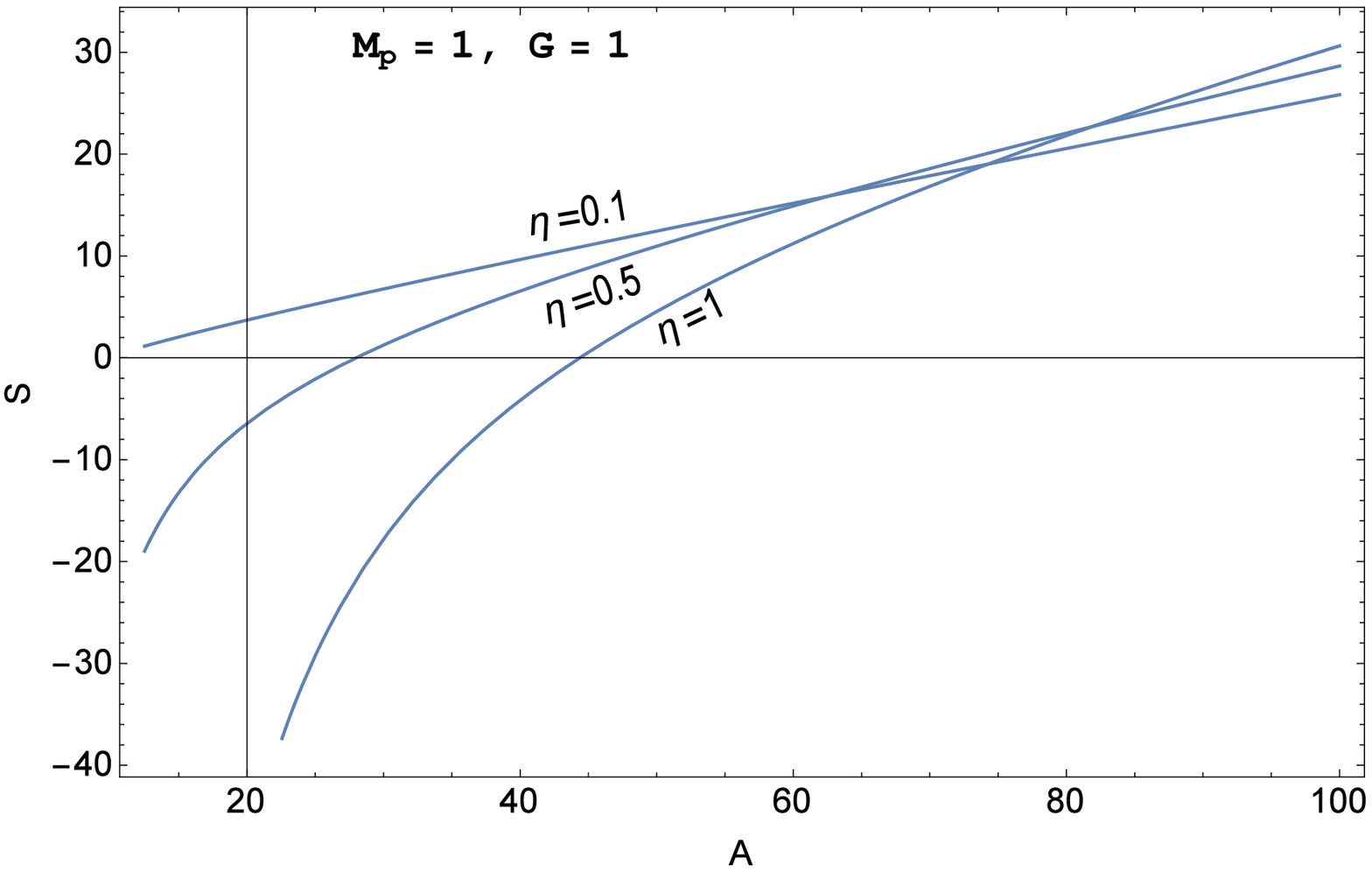}
\end{center}
\end{figure}\\ \\
For $n=1$ and $2$ curves of $S$ for $\eta = 1,0.5$ and $0.1$ are been given in figure 3.2 and 3.3. It shows that more the $\eta$, more is the A to start the black hole's journey. The average slope of the curve is also higher if $\eta$ is higher.\\
$\frac{dS}{dA} \sim \frac{dS}{dM} = \frac{1}{T}$. From this  we can say that if $\eta$ increases then $\frac{dS}{dA}$ increases and hence temperature of the system decreases. So here $\eta$ is somehow representative of the less temperature.\\
\newpage
\section{Thermodynamics of rainbow gravity inspired Reissner–Nordstrom black hole}
The Reissner-Nordstrom black hole metric under the effect of rainbow gravity is given by \\
\begin{equation}
ds^2=\frac{1}{f^2\left(\frac{E}{E_{p}}\right)}\left[1-\frac{2M}{r}+ \frac{Q^2}{r^2}\right] dt^2 + \frac{1}{g^2\left(\frac{E}{E_{p}}\right)}\left[1-\frac{2M}{r}+ \frac{Q^2}{r^2}\right]^{-1} dr^2 + \frac{r^2}{g^2\left(\frac{E}{E_{p}}\right)}d\Omega^2 \\
\end{equation}
The surface gravity follows from (\ref{Surface_gravity_rainbow}) as \\
\begin{equation}
\kappa = \frac{g\left(\frac{E}{E_{p}}\right)}{f\left(\frac{E}{E_{p}}\right)} \left(\frac{M}{R^2_{N}}-\frac{Q^2}{R^3_{N}}\right)
\end{equation}
where $R_{N}=M+\sqrt{M^2-Q^2}$ is the radius of the event horizon of the RN BH under rainbow gravity. The Hawking temperature is given by \\
\begin{equation}\label{Temperature_RN}
T=\frac{1}{2\pi}\sqrt{1-\frac{\eta}{(E_{p})^n}\frac{1}{R^n_{N}}}\left(\frac{M}{R^2_{N}}-\frac{Q^2}{R^3_{N}}\right)
\end{equation}
For getting the above relation we have put $E=\frac{1}{R_{N}}$. \\
\begin{figure}[h]
~~~~~~~~~~Fig.4a~~~~\hspace{3cm}~~~~~~~~~~~~~Fig.4b~~~~~~~~\hspace{3.2cm}~~~Fig.4c\\
\includegraphics[height=1.2in, width=2.3in]{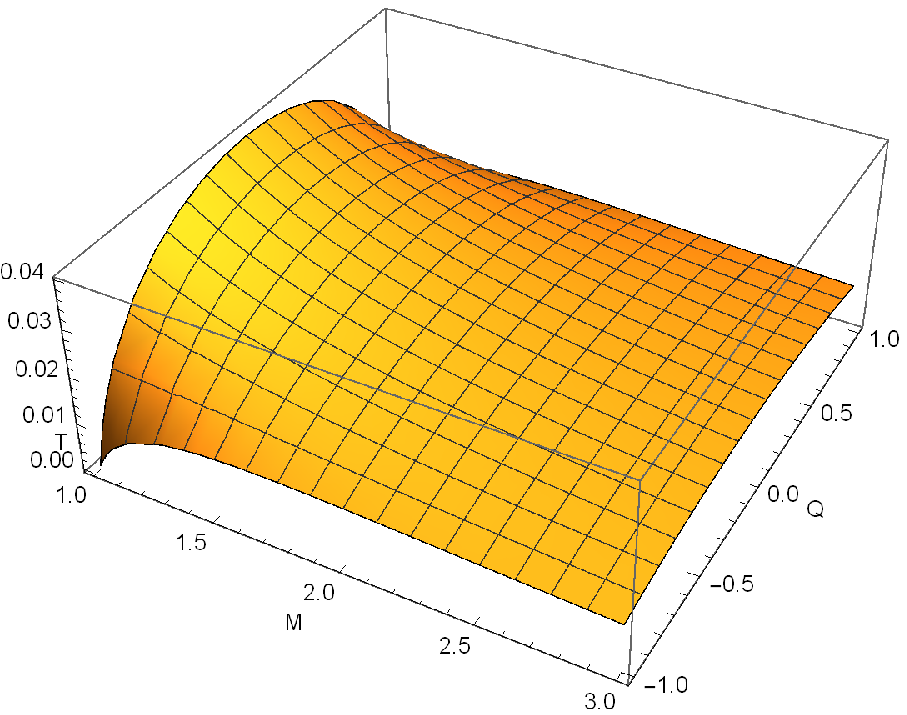}~~
\includegraphics[height=1.2in, width=2.3in]{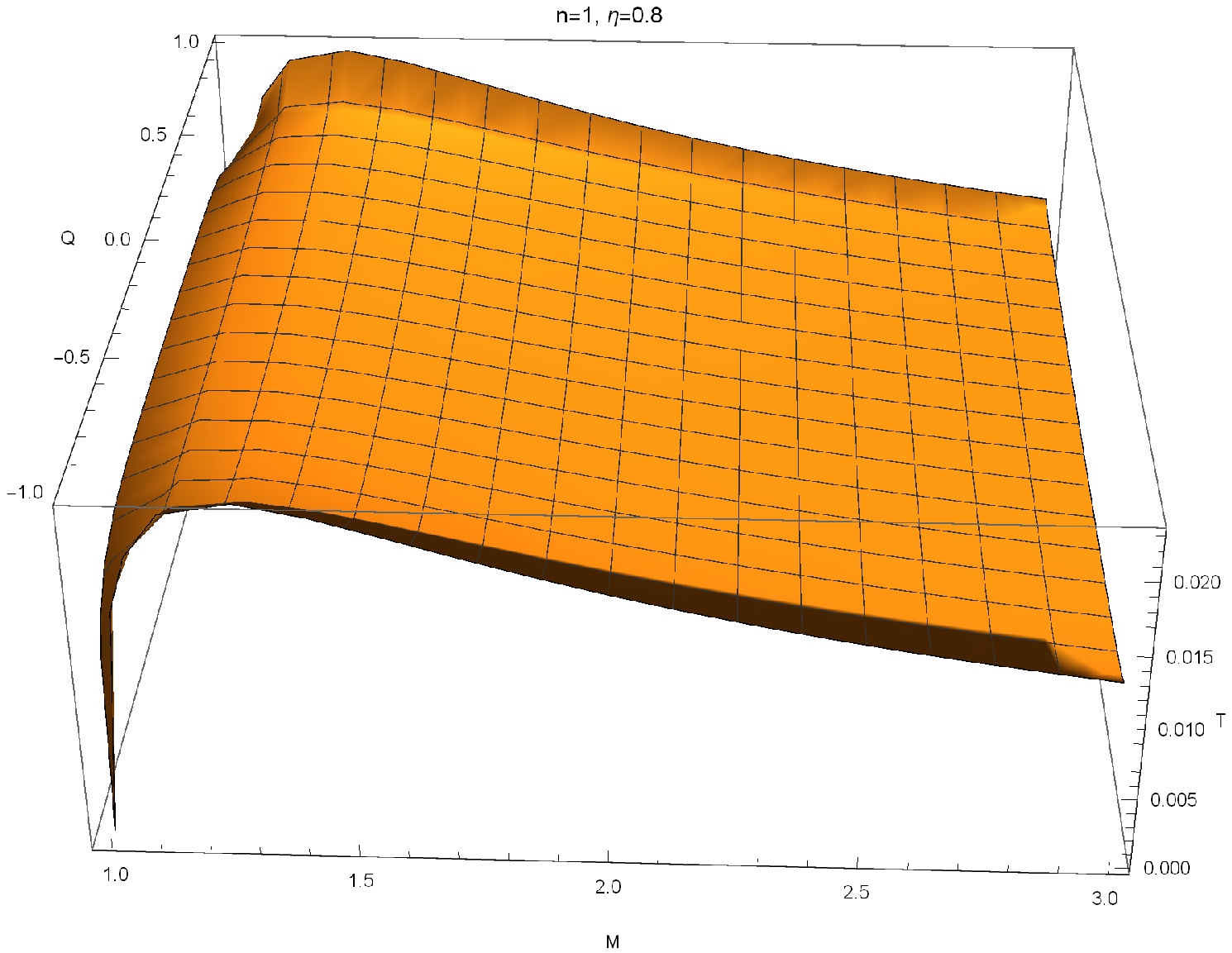}~~
\includegraphics[height=1.2in, width=2.3in]{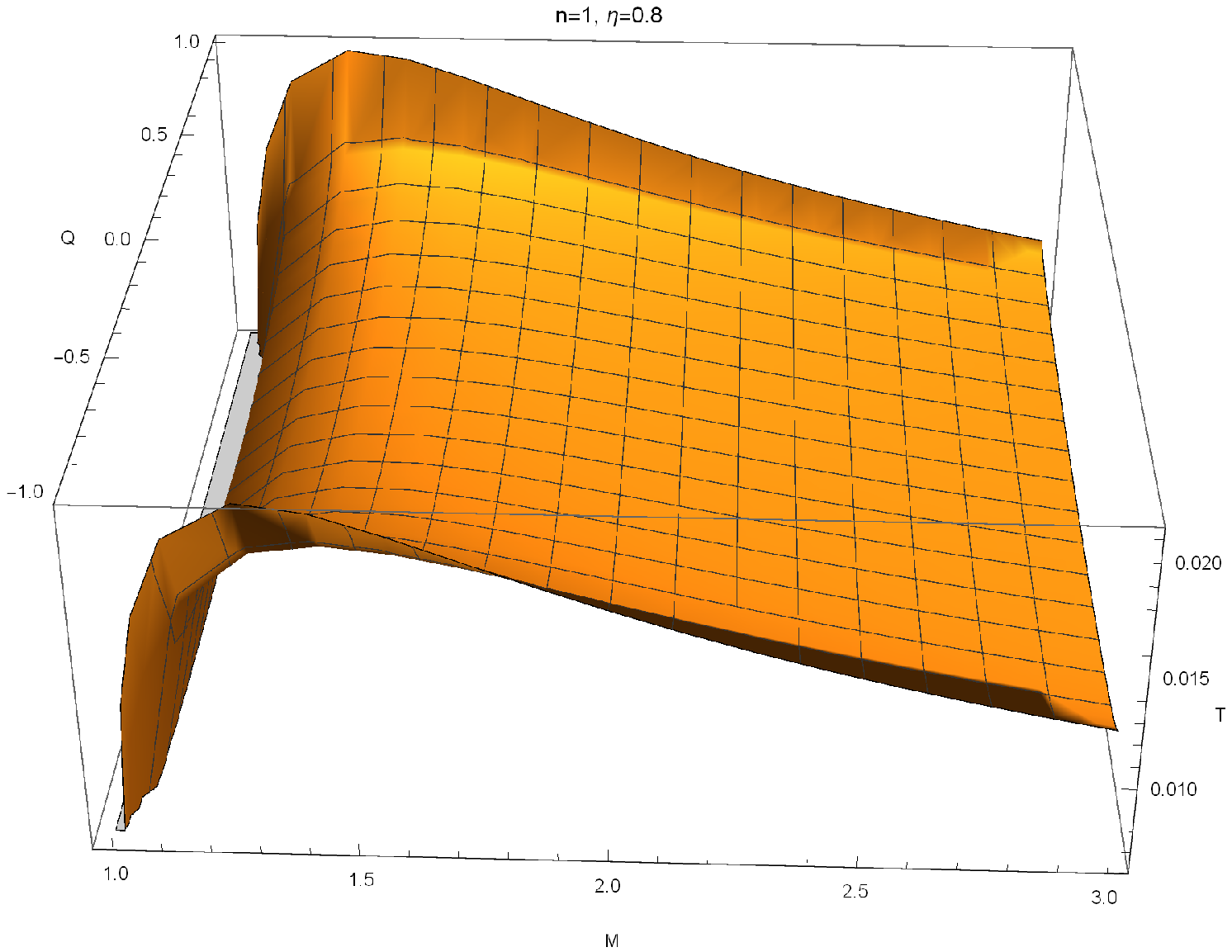}~~\\
\vspace{.1cm}Fig $4a$. represents $T$ vs $M$ and $Q$ in Einstein gravity.\\
Fig $4b$. represents $T$ vs $M$ and $Q$ in Rainbow gravity with $\eta=0.8$.\\
Fig $4c$. represents $T$ vs $M$ and $Q$ in Rainbow gravity with $\eta=1$.
\end{figure}\\
Every thermodynamic quantity of RN BH primarily consists of two variables $M$ and $Q$. For Einstein gravity if our BH comprises of low charge then temperature decreases with increasing mass. But as we increase $|Q|$ the temperature curve is broken into two phases, first increasing to a particular local maxima and then decreasing. We can speculate that even if in Einstein gravity a phase transition may occur for highly charged RN BH. But if we look for Rainbow gravity effect on temperature  (Fig.4c) we can see that, whatever be the value of $|Q|$, we will always get two distinct phases : firstly increasing and then decreasing after attaining a local maxima. When charge is high the local maxima is also higher.\\
Since the temperature must be real quantity, so we must have  \\
\begin{equation}
1-\frac{\eta}{(E_{p})^n}\frac{1}{R^n_{N}} \geq 0
\end{equation}
hence the critical mass $M_{cr}$ of the RN BH under the effect of rainbow gravity is given by  \\
\begin{equation}
M_{cr}=\frac{1}{2}\left[\frac{Q^2}{\left(\frac{\eta}{E^n_{p}}\right)^\frac{1}{n}}+\left(\frac{\eta}{E^n_{p}}\right)^\frac{1}{n}\right]
\end{equation}
From (\ref{Temperature_RN}) we get  \\
\begin{center}
$\frac{dT}{dM}=\frac{(E_{p} R_{N})^{-n} \left[Q^2 \left(6 (E_{p} R_{N})^n-\eta  (n+6)\right)+R^2_{N} \left(\eta  (n+2)-2 (E_{p} R_{N})^n\right)\right]}{4 \pi  R^2_{N} \left(R^2_{N}-Q^2\right) \sqrt{1-\eta  (E_{p} R_{N})^{-n}}}$
\end{center}
The heat capacity C of this BH is given by \\
\begin{equation}
C=\frac{dM}{dT}= \frac{4 \pi  R^2_{N} (Q^2-R^2_{N}) (E_{p} R_{N})^n \sqrt{1-\eta  (E_{p} R_{N})^{-n}}}{2 \left(R^2_{N}-3 Q^2\right) (E_{p} R_{N})^n+\eta  \left[(n+6) Q^2-(n+2) R^2_{N}\right]}
\end{equation}
\begin{figure}[h]
\begin{center}
Fig.5\\
\includegraphics[height=2.9in, width=3.6in]{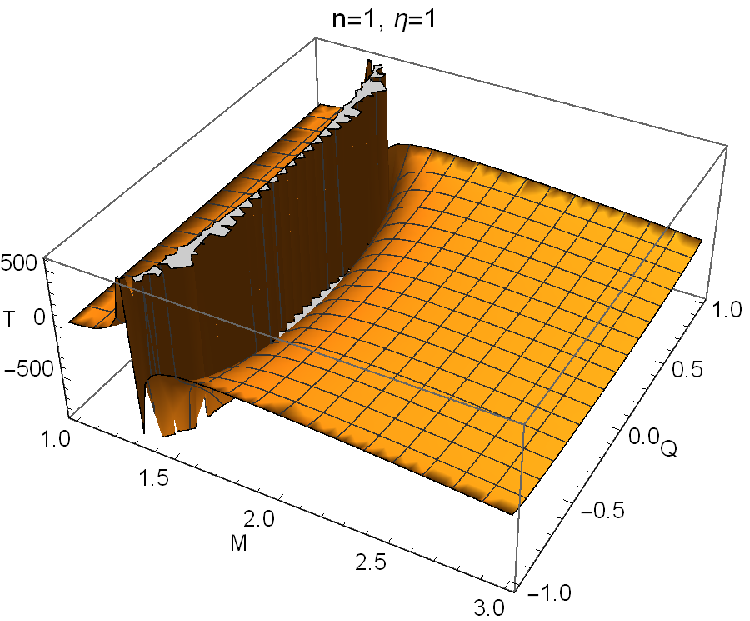}\\
Fig. $5$ represents $C$ vs $M$ and $Q$ for $n=1$. This shows that there must be a phase transition turning the BH from stable to unstable phase. Larger the $|Q|$, larger the $M$ where the transition to be occurred.
\end{center}
\end{figure}\\ \\
As before by putting $C = 0$ we get the remnant mass $M_{rem}$ given by  \\
\begin{equation}
M_{rem}= \frac{1}{2}\left[\frac{Q^2}{\left(\frac{\eta}{E^n_{p}}\right)^\frac{1}{n}}+\left(\frac{\eta}{E^n_{p}}\right)^\frac{1}{n}\right]
\end{equation}
this is found to be the same as the critical mass. But as before this is actually the starting mass.\\
The entropy of this black hole is now computed keeping terms upto $O(\eta^4)$ and assuming that $n\geq 3$ leads to:\\
$
S=2\pi \mathlarger{ \int}\frac{dM}{\sqrt{1-\frac{\eta}{(E_{p})^n}\frac{1}{R^n_{N}}}\left(\frac{M}{R^2_{N}}-\frac{Q^2}{R^3_{N}}\right)}$ \\
$
=2\pi \mathlarger{\int} \left[R_{N}+ \frac{\eta}{E^n_{p}}\frac{1}{R^{n-1}_{N}}+ \frac{3\eta^2}{8E^{2n}_{p}}\frac{1}{R^{2n-1}_{N}} +\frac{5\eta^3}{16E^{3n}_{p}}\frac{1}{R^{3n-1}_{N}}+ \frac{35\eta^4}{128E^{4n}_{p}}\frac{1}{R^{4n-1}_{N}}\right]dR_{N}
$ \\
$
=2\pi\left[\frac{R^2_{N}}{2}+\frac{\eta}{(2-n)E^n_{p}R^{n-2}_{N}}+\frac{3\eta^2}{8(2-2n)E^{2n}_{p}R^{2n-2}_{N}}+\frac{5\eta^3}{16(2-3n)E^{3n}_{p}R^{3n-2}_{N}}+\frac{35\eta^4}{128(2-4n)E^{4n}_{p}R^{4n-2}_{N}}\right]
$\\
\begin{equation}
=S_{BH}+\frac{2\pi^{\frac{n}{2}}\eta}{(2-n)E^n_{p}S^{\frac{n}{2}-1}_{BH}}+ \frac{3\pi^n\eta^2}{8(1-n)E^{2n}_{p}S^{n-1}_{BH}}+ \frac{5\pi^{\frac{3n}{2}}\eta^3}{8(2-3n)E^{3n}_{p}S^{\frac{3n}{2}-1}_{BH}}+ \frac{35\pi^{2n}\eta^4}{128(1-2n)E^{4n}_{p}S^{2n-1}_{BH}}
\end{equation}
where $S_{BH} = \pi R^2_{N}$ is the semi-classical Bekenstein-Hawking entropy of the RN BH under the rainbow gravity.\\
$
=> \mathlarger{ S =\left( \frac{A}{4} \right)+\frac{2\pi^{\frac{n}{2}}\eta}{(2-n)E^n_{p}\left( \frac{A}{4} \right)^{\frac{n}{2}-1}}+ \frac{3\pi^n\eta^2}{8(1-n)E^{2n}_{p}\left( \frac{A}{4} \right)^{n-1}}+ \frac{5\pi^{\frac{3n}{2}}\eta^3}{8(2-3n)E^{3n}_{p}\left( \frac{A}{4} \right)^{\frac{3n}{2}-1}}}$
\begin{equation}
+\frac{35\pi^{2n}\eta^4}{128(1-2n)E^{4n}_{p}\left( \frac{A}{4} \right)^{2n-1}}
\end{equation}\\
\begin{figure}[h]
\begin{center}
Fig.6\\
\includegraphics[height=2.5in, width=3.2in]{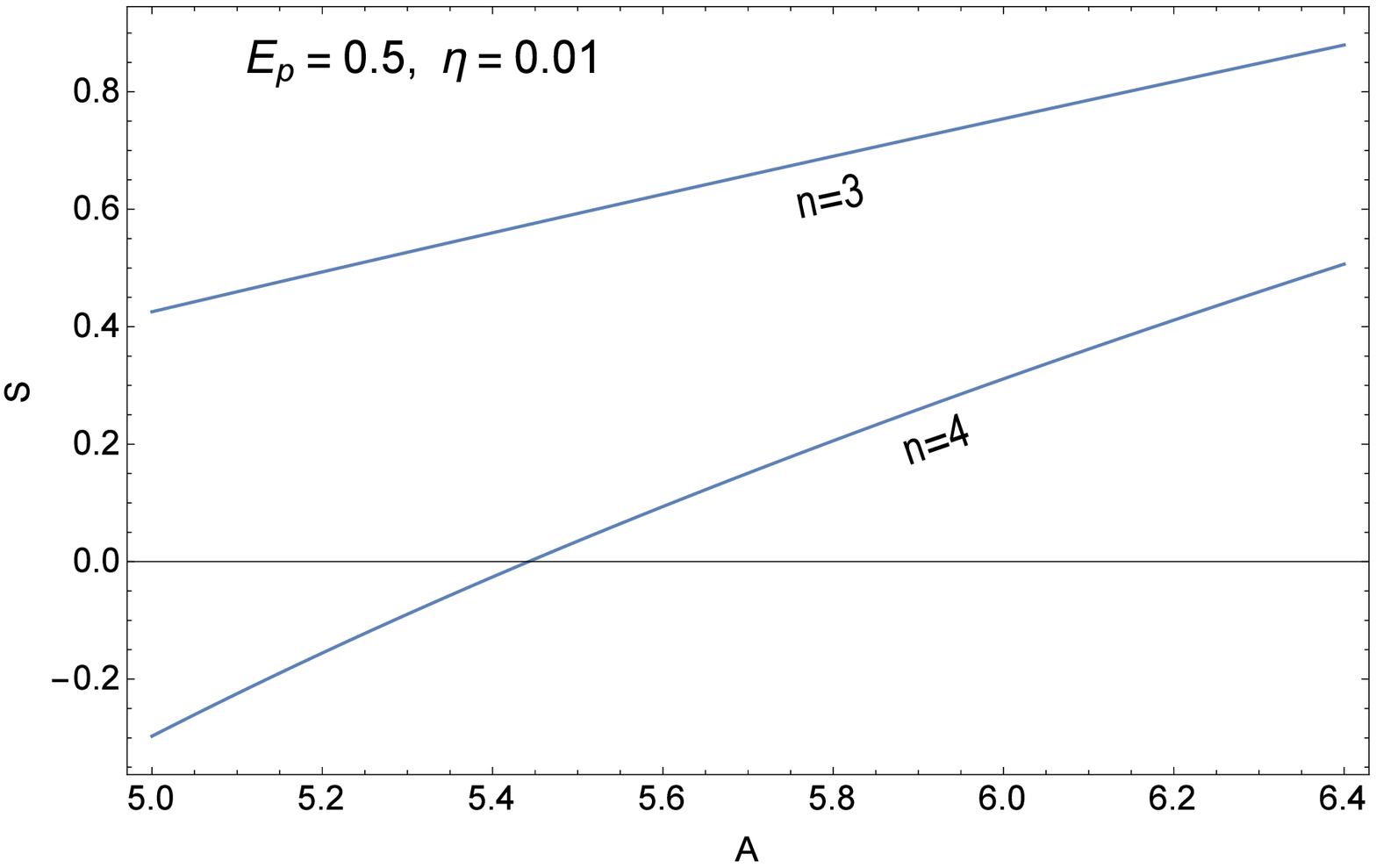}\\
Fig.6 depicts the S vs A curve. S is increasing with A. If we increase n, the curve is shifted downwards, i.e., more the 'n' less the entropy.
\end{center}
\end{figure}\\
For n=1 we have : \\ \\
$
S=2\pi \mathlarger{ \int \left[R_{N}+ \frac{\eta}{E_{p}}+ \frac{3\eta^2}{8E^{2}_{p}}\frac{1}{R_{N}} +\frac{5\eta^3}{16E^{3}_{p}}\frac{1}{R^{2}_{N}}+ \frac{35\eta^4}{128E^{4}_{p}}\frac{1}{R^{3}_{N}}\right]dR_{N}}
$ \\ \\
$
=\mathlarger{S_{BH}+ \frac{2\eta \sqrt{\pi}\sqrt{S_{BH}}}{E_{p}}+\frac{3\eta^2 \pi \ ln\left(\frac{S_{BH}}{\pi}\right)}{8E^2_{p}}-\frac{5\eta^3 \pi^{\frac{3}{2}}}{8E^{3}_{p}\sqrt{S_{BH}}}-\frac{35\eta^4 \pi^2}{128E^4_{p}S_{BH}}}
$ \\ \\
\begin{equation}
=> S=\left(\frac{A}{4}\right)+ \frac{2\eta \sqrt{\pi}\sqrt{\left(\frac{A}{4}\right)}}{E_{p}}+\frac{3\eta^2 \pi \ ln\left(\frac{\left(\frac{A}{4}\right)}{\pi}\right)}{8E^2_{p}}-\frac{5\eta^3 \pi^{\frac{3}{2}}}{8E^{3}_{p}\sqrt{\left(\frac{A}{4}\right)}}-\frac{35\eta^4 \pi^2}{128E^4_{p}\left(\frac{A}{4}\right)}
\end{equation} 
For n=2 we have : \\ \\
$
S=2\pi \mathlarger{\int \left[R_{N}+ \frac{\eta}{E^2_{p}}\frac{1}{R_{N}}+ \frac{3\eta^2}{8E^{4}_{p}}\frac{1}{R^{3}_{N}} +\frac{5\eta^3}{16E^{6}_{p}}\frac{1}{R^{5}_{N}}+ \frac{35\eta^4}{128E^{8}_{p}}\frac{1}{R^{7}_{N}}\right]dR_{N}}
$ \\ \\
$\mathlarger{=S_{BH}+\frac{\eta \pi \ ln\left(\frac{S_{BH}}{\pi}\right)}{E^2_{p}}-\frac{3\eta^2 \pi^2}{8 E^4_{p}S_{BH}}- \frac{5\eta^3 \pi^3}{32E^6_{p}S^2_{BH}}- \frac{35\eta^4 \pi^4}{384 E^8_{p}S^3_{BH}}}
$
\begin{equation}
=> S = \left(\frac{A}{4}\right)+\frac{\eta \pi \ ln\left(\frac{\left(\frac{A}{4}\right)}{\pi}\right)}{E^2_{p}}-\frac{3\eta^2 \pi^2}{8 E^4_{p}\left(\frac{A}{4}\right)}- \frac{5\eta^3 \pi^3}{32E^6_{p}\left(\frac{A}{4}\right)^2}- \frac{35\eta^4 \pi^4}{384 E^8_{p}\left(\frac{A}{4}\right)^3}
\end{equation}
\begin{figure}[h]
\begin{center}
Fig.7.1~~~~~~~~\hspace{3.5cm}~~~~~~~~~~~~~~~~~~Fig.7.2\\
\includegraphics[height=2.5in, width=3.2in]{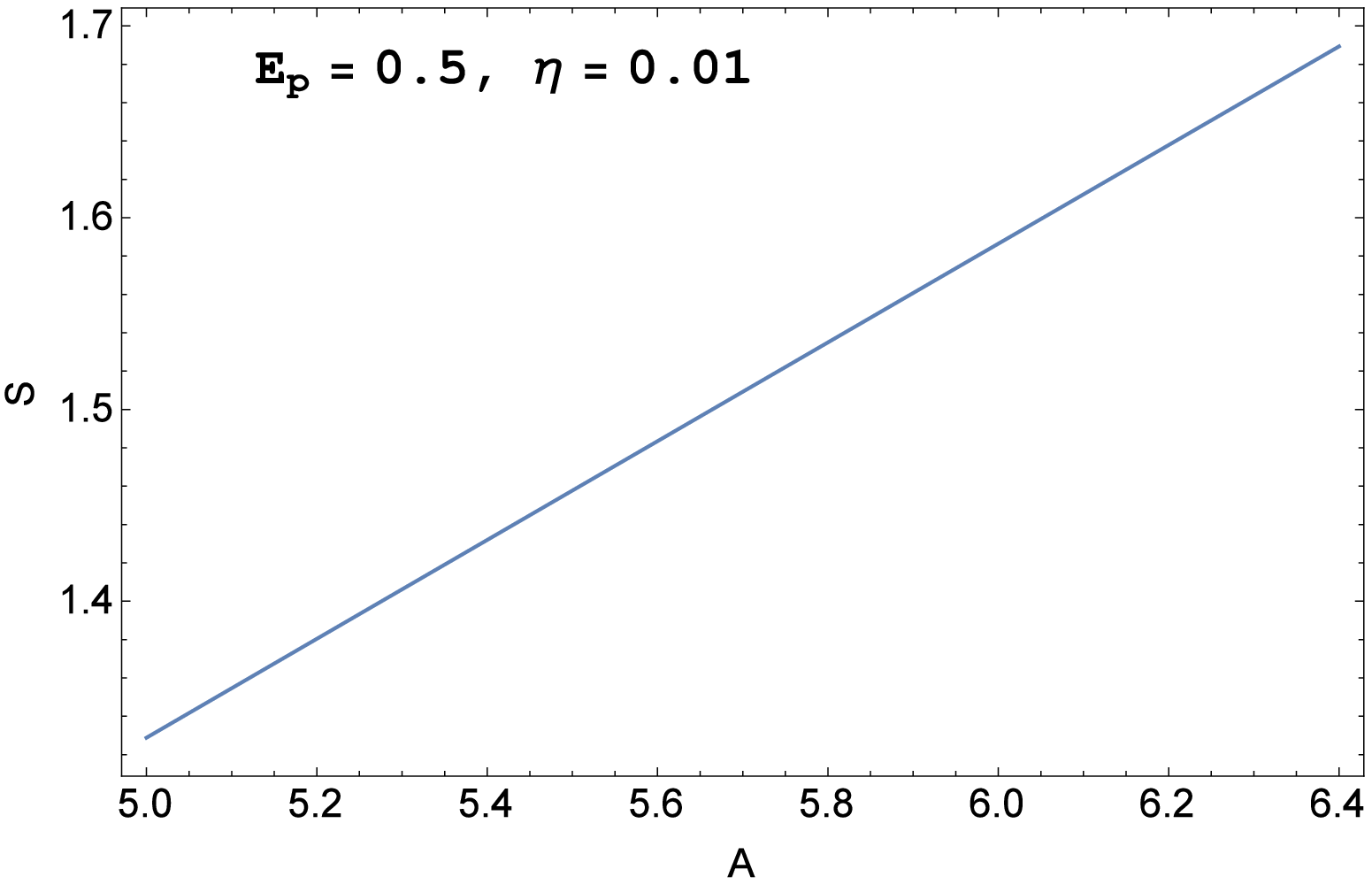}
\includegraphics[height=2.5in, width=3.2in]{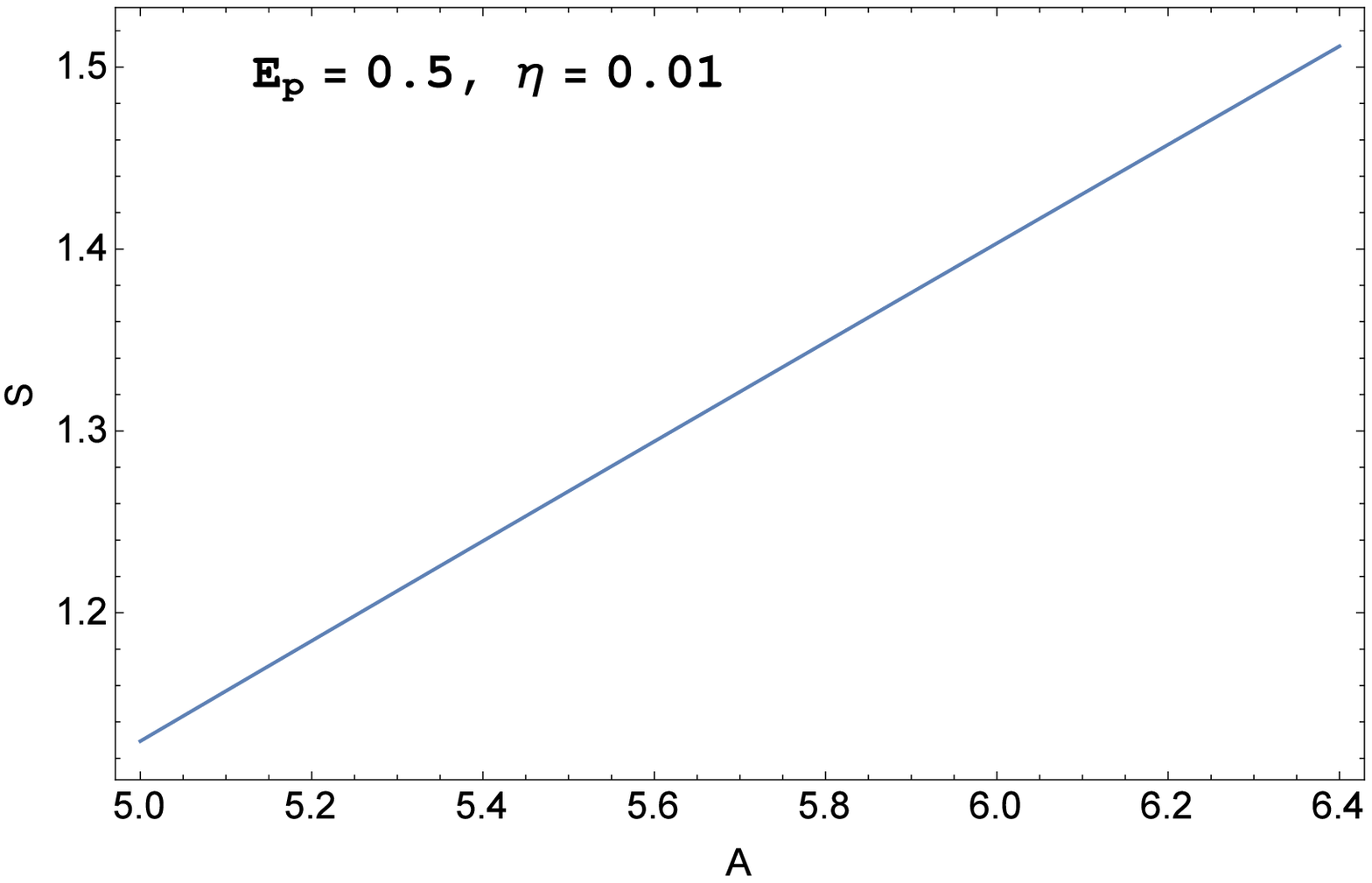}\\
Figures 7.1 and 7.2 both show that S increases with A.
\end{center}
\end{figure}
\pagebreak \\
\section{Thermodynamics of rainbow gravity inspired Reissner Nordstrom de Sitter black hole surrounded by quintessence} 
In this section we want to study the thermodynamic properties of Reissner Nordstrom de Sitter black hole. The metric of RN-de Sitter inspired by rainbow gravity is given by : \\
\begin{equation}
ds^2=\frac{1}{f^2\left(\frac{E}{E_{p}}\right)}\left[1-\frac{r_{g}}{r}-\sum_{n}{\left(\frac{r_{n}}{r}\right)^{3w_{q}+1}}\right]dt^2 -\frac{1}{g^2\left(\frac{E}{E_{p}}\right)}\frac{dr^2}{\left[1-\frac{r_{g}}{r}-\sum_{n}{\left(\frac{r_{n}}{r}\right)^{3w_{q}+1}}\right]}- \frac{r^2}{g^2\left(\frac{E}{E_{p}}\right)}d\Omega^2 \\
\end{equation}
where $r_{g}=2M$, M is the mass of the black hole, $r_{n}-s$ are the dimensional normalisation constants and $w_{q}$ are the quintessential state parameters.\\
The work of Kiselev has provided a particular solution for the Reissner-Nordstrom-de Sitter black hole surrounded by quintessence as : \\
$g_{tt}^{QdS}= g_{tt}= \frac{1}{f^2\left(\frac{E}{E_{p}}\right)} \left[1 -\frac{r_{g}}{r}+\frac{Q^2}{r^2}-\frac{r^2}{a^2}-\left(\frac{r_{q}}{r}\right)^{3w_{q}+1}\right]$ \\
This solution in its more particular case turns to meaningful limits with no charge $(Q=0)$.\\
The surface gravity relation is defined by : \\
$k=\lim_{r \to R_{s}} \sqrt{-\frac{1}{4}g^{rr}g^{tt}(g_{tt,r})^2}$ \\
where $R_{s}=2GM$ is the Schwarzschild radius. \\
Using the surface gravity relation we get :
\begin{equation}
k=\frac{g(\frac{E}{E_{p}})}{f(\frac{E}{E_{p}})}\frac{1}{4MG}\left[\frac{1}{G}-\frac{8M^2G^2}{a^2}+(3w_{q}+1)\left(\frac{r_{q}}{2G}\right)^{3w_{q}+1}\frac{1}{M^{3w_{q}+1}}\right]
\end{equation}
Using (2) we can obtain :
\begin{equation}
T=\frac{1}{8\pi G}\sqrt{\frac{1}{M^2}-\frac{\eta}{(2GE_{p})^n}\frac{1}{M^{n+2}}}\left[ \frac{1}{G}-\frac{8M^2G^2}{a^2}+(3w_{q}+1)\left(\frac{r_{q}}{2G}\right)^{3w_{q}+1}\frac{1}{M^{3w_{q}+1}} \right]
\end{equation}
In the above expression we have set $E=\frac{1}{2GM}.$ \\
\begin{figure}[h]
\begin{center}
Fig.9\\
\includegraphics[height=2.5in, width=3.2in]{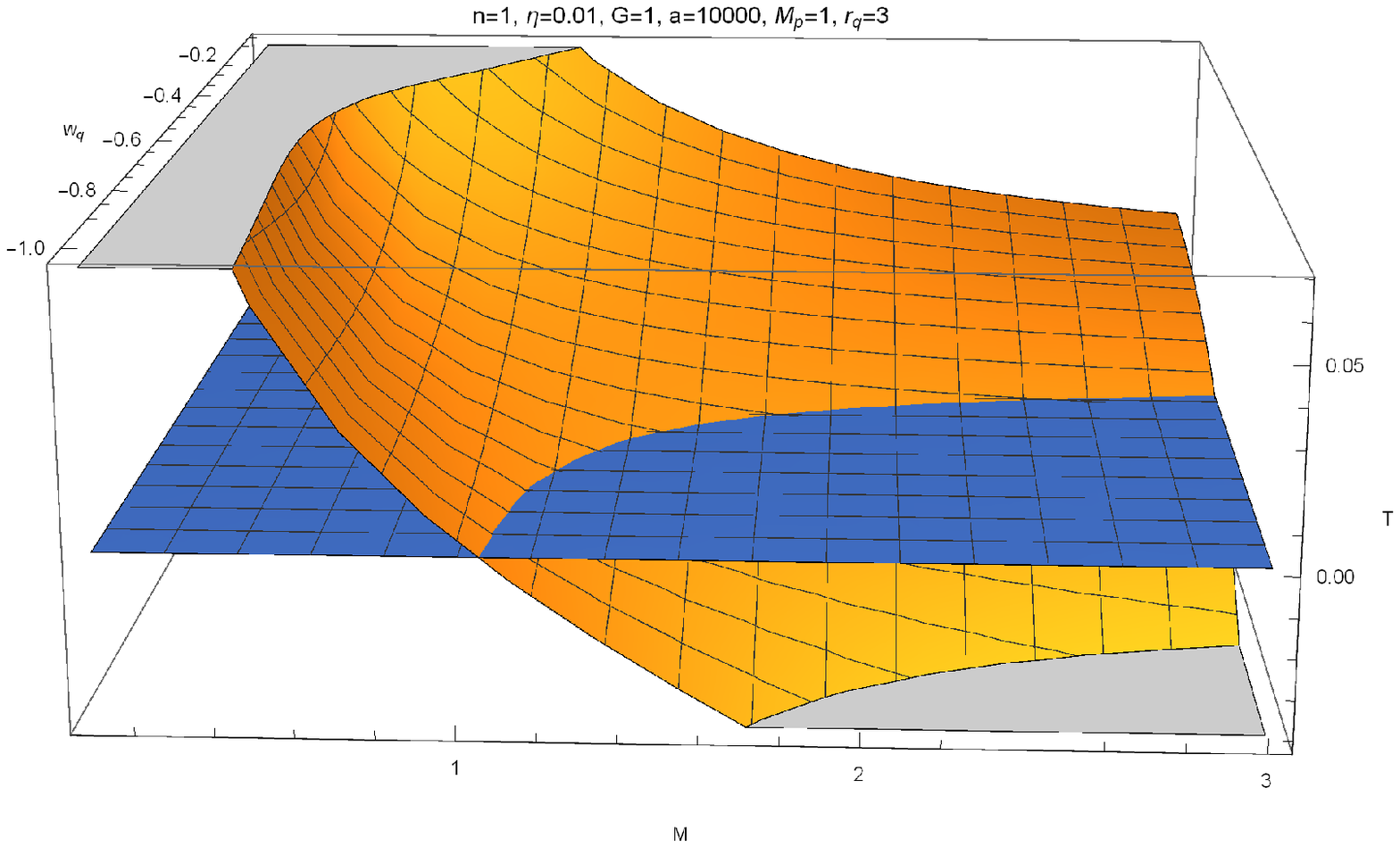}\\
Fig.9 represents T vs M and $w_q$ graph.
\end{center}
\end{figure}\\
This equation gives us a relation between the temperature and the mass. Since the temperature has to be a real quantity, we obtain the following condition :
\begin{equation}
\frac{1}{M^2}-\frac{\eta}{(2GE_{p})^n}\frac{1}{M^{n+2}}\geq 0 .
\end{equation}
The above condition readily leads to the existence of a critical mass $(M_{cr})$ below which the
temperature becomes a complex quantity. This critical mass is given by :
\begin{equation}
M_{cr}=\frac{\eta^{\frac{1}{n}}}{2GE_{p}}=\eta^{\frac{1}{n}}M_{p} .
\end{equation}
where we have taken $E_{p}=\frac{1}{2GM_{p}} \ $ .
From (18) we get :\\
$\frac{dT}{dM}=\frac{\left(\frac{(n+2)\eta}{(2GE_{p})^n}\frac{1}{M^{n+3}}-\frac{2}{M^3}\right)}{16\pi G\sqrt{\frac{1}{M^2}-\frac{\eta}{(2GE_{p})^n}\frac{1}{M^{n+2}}}}\left[ \frac{1}{G}-\frac{8M^2G^2}{a^2}+(3w_{q}+1)\left(\frac{r_{q}}{2G}\right)^{3w_{q}+1}\frac{1}{M^{3w_{q}+1}} \right] + $ \\ \\
$ \\
\frac{1}{8\pi G}\sqrt{\frac{1}{M^2}-\frac{\eta}{(2GE_{p})^n}\frac{1}{M^{n+2}}}\left[ -\frac{16MG^2}{a^2}- (3w_{q}+1)^2\left(\frac{r_{q}}{2G}\right)^{3w_{q}+1}\frac{1}{M^{3w_{q}+2}} \right]$ \\ \\
The heat capacity reads : \\
$C=\frac{dM}{dT} $\\
$= 16\pi G\sqrt{\frac{1}{M^2}-\frac{\eta}{(2GE_{p})^n}\frac{1}{M^{n+2}}}\left[\left(\frac{(n+2)\eta}{(2GE_{p})^n}\frac{1}{M^{n+3}}-\frac{2}{M^3}\right)\left\{ \frac{1}{G}-\frac{8M^2G^2}{a^2}+(3w_{q}+1)\left(\frac{r_{q}}{2G}\right)^{3w_{q}+1}\frac{1}{M^{3w_{q}+1}} \right\}\right.$

$\left.+2\left(\frac{1}{M^2}-\frac{\eta}{(2GE_{p})^n}\frac{1}{M^{n+2}}\right)\left\{ -\frac{16MG^2}{a^2}- (3w_{q}+1)^2\left(\frac{r_{q}}{2G}\right)^{3w_{q}+1}\frac{1}{M^{3w_{q}+2}} \right\}\right]^{-1} $ \\ \\
The remnant mass $M_{rem}$(where the black hole stops evaporating) can be obtained by setting
$C=0$. This yields :\\
\begin{equation}
M_{rem}=\frac{\eta^{\frac{1}{n}}}{2GE_{p}}=\eta^{\frac{1}{n}}M_{p} .
\end{equation}
Thus we can see that the remnant mass of the black hole is equal to its critical
mass.
The entropy can be calculated by using the heat capacity of this black hole given by relation (11). \\
Substituting equation (18) in equation (11) we get :
\begin{equation}
 S=\int\frac{8\pi G dM}{\sqrt{\frac{1}{M^2}-\frac{\eta}{(2GE_{p})^n}\frac{1}{M^{n+2}}}\left[ \frac{1}{G}-\frac{8M^2G^2}{a^2}+(3w_{q}+1)\left(\frac{r_{q}}{2G}\right)^{3w_{q}+1}\frac{1}{M^{3w_{q}+1}} \right]}
 \end{equation} \\
We take values of $w_{q}= -\frac{1}{3}, -\frac{2}{3}$ and -1 and carrying out a binomial expansion keeping terms upto $O(\eta^{4})$ leads to three different cases respectively. \\
 $
{\large \textbf{Case-1} : w_{q}=-\frac{1}{3}}$ \\ \\
$ \mathlarger{ S=8\pi a^2 G^2 \int \ \frac{\left[ M + \frac{\eta}{2(2GE_{p})^n}\frac{1}{M^{n-1}} + \frac{3\eta^{2}}{8(2GE_{p})^{2n}}\frac{1}{M^{2n-1}} + \frac{5\eta^{3}}{16(2GE_{p})^{3n}}\frac{1}{M^{3n-1}}+\frac{35\eta^{4}}{128(2GE_{p})^{4n}}\frac{1}{M^{4n-1}}\right]dM}{\left[ a^2- 8M^2G^3 \right]}}$ \\ \\ 
 {\large\textbf{Case-1(a)}} : n=1 \\
$
S= -\left[8\pi a^2 G^2\frac{ln(a^2-8G^3 M^2)}{16G^3}\right] + \left[8\pi a^2 G^2\frac{\eta \ tanh^{-1}\left(\frac{2\sqrt{2}G^{\frac{3}{2}}M}{a}\right)}{4\sqrt{2}a(2 G E_{p})G^{\frac{3}{2}}}\right] + \left[8\pi a^2 G^2\frac{3\eta^2 \ ln\left(\frac{M^2}{a^2-8G^3 M^2}\right)}{16a^2 (2G E_{p})^2}\right]$\\
$ - \left[8\pi a^2 G^2\frac{5\eta^3\left[ a-2\sqrt{2}G^{\frac{3}{2}}M tanh^{-1}\left(\frac{2\sqrt{2}G^{\frac{3}{2}}M}{a}\right)\right]}{16a^3 (2G E_{p})^3 M}\right] - \left[8\pi a^2 G^2\frac{35\eta^4\left[ a^2 + 8G^3M^2 ln\left( \frac{a^2-8G^3M^2}{M^2} \right) \right]}{256a^4(2GE_{p})^4M^2} \right]
$ \\ \\
\resizebox{.7 \textwidth}{!} 
{
$
=\frac{5 \pi  \eta ^3 G^2 M_p^3}{32}  \left(\frac{8}{a^2-8 G^3 M^2}-\frac{7 \eta  M_p}{M^2}\right)+ \pi \eta M_{p}a \sqrt{2G}\ tanh^{-1}\left(\frac{2\sqrt{2}G^{\frac{3}{2}}M}{a}\right) 
$
}\\ \\
\resizebox{.8 \textwidth}{!} 
{
$
\hspace{3.5cm}+ \frac{1}{4a^2} ln \left(\frac{M^2}{a^2-8 G^3 M^2}\right) \left(\pi  \eta ^2 G^2 M_p^2\right) \left(6 a^2+35 \eta ^2 G^3 M_p^2+5 \eta  M_p\right)
$
}\\
$
=\frac{5 \pi  \eta ^3 G^2 M_p^3}{32}  \left(\frac{8\pi}{\pi a^2-2 G^3 M^2_{p}S_{BH}}-\frac{28 \eta \pi}{M_{p}S_{BH}}\right)+ \pi \eta M_{p}a \sqrt{2G}\ tanh^{-1}\left(\frac{\sqrt{2S_{BH}}G^{\frac{3}{2}}M_{p}}{a\sqrt{\pi}}\right)
$
\begin{equation}
\resizebox{.67 \textwidth}{!} 
{
$
+ \frac{1}{4a^2} ln \left(\frac{M^2_{p}S_{BH}}{4\pi a^2-8 G^3 M^2_{p}S_{BH}}\right) \left(\pi  \eta ^2 G^2 M_p^2\right) \left(6 a^2+35 \eta ^2 G^3 M_p^2+5 \eta  M_p\right)$
}
\end{equation}
$
=\frac{5 \pi  \eta ^3 G^2 M_p^3}{32}  \left(\frac{8\pi}{\pi a^2-2 G^3 M^2_{p}\left(\frac{A}{4}\right)}-\frac{28 \eta \pi}{M_{p}\left(\frac{A}{4}\right)}\right)+ \pi \eta M_{p}a \sqrt{2G}\ tanh^{-1}\left(\frac{\sqrt{\left(\frac{A}{2}\right)}G^{\frac{3}{2}}M_{p}}{a\sqrt{\pi}}\right) $\\ 
\begin{equation}
\resizebox{.67 \textwidth}{!} 
{
$
+ \frac{1}{4a^2} ln \left(\frac{M^2_{p}\left(\frac{A}{4}\right)}{4\pi a^2-8 G^3 M^2_{p}\left(\frac{A}{4}\right)}\right) \left(\pi  \eta ^2 G^2 M_p^2\right) \left(6 a^2+35 \eta ^2 G^3 M_p^2+5 \eta  M_p\right)$
}
\end{equation}\\
{\large\textbf{Case-1(b)}} : n=2 \\ \\
$
S= -\frac{\pi a^2}{2G}\ ln(a^2-8G^3 M^2) +\ 2\eta \ G^2 M^2_{p} \pi \ ln\left( \frac{M^2}{a^2-8G^3M^2}\right) + \frac{\left(12 \pi  \eta ^2 G^5 M_p^4\right)}{a^2}\ ln \left(\frac{M^2}{a^2-8 G^3 M^2}\right)$ \\ \\
$
- \frac{3 \pi  \eta ^2 G^2 M_p^4}{2 M^2} -\frac{10 \pi  \eta ^3 G^5 M_p^6}{a^2 M^2}+\frac{ \left(80 \pi  \eta ^3 G^8 M_p^6\right)}{a^4} ln \left(\frac{M^2}{a^2-8 G^3 M^2}\right)-\frac{5 \pi  \eta ^3 G^2 M_p^6}{8 M^4}
$ \\ \\
$
= -\frac{35 \pi  M_p^8 \eta ^4  G^2}{96 M^6} -\frac{5 \pi  M_p^6 \eta ^3 G^2 \left(a^2+7 M_p^2 \eta  G^3\right)}{8 a^2 M^4}-\frac{\pi  M_p^4 \eta ^2 G^2 \left(3 a^4+20 a^2 M_p^2 \eta  G^3+140 M_p^4 \eta ^2 G^6\right)}{2 a^4 M^2}$ \\  \\
$
+\frac{ln (M) \left(4 \pi  M_p^2 \eta  G^2\right) \left(a^6+6 a^4 M_p^2 \eta  G^3+40 a^2 M_p^4 \eta ^2 G^6+280 M_p^6 \eta ^3 G^9\right)}{a^6} $ \\ \\
$-\frac{\pi  \left(a^8+4 a^6 M_p^2 \eta  G^3+24 a^4 M_p^4 \eta ^2 G^6+160 a^2 M_p^6 \eta ^3 G^9+1120 M_p^8 \eta ^4 G^{12}\right) ln \left(a^2-8 G^3 M^2\right)}{2 a^6 G} $ \\ \\
$
= -\frac{70 \pi^4  M_p^2 \eta ^4 G^2}{3S^3_{BH}} -\frac{10 \pi^3  M_p^2 \eta^3 G^2 \left(a^2+7 M_p^2 \eta  G^3\right)}{ a^2 S^2_{BH}}- \frac{2\pi^2  M_p^2 \eta^2 G^2 \left(3 a^4+20 a^2 M_p^2 \eta  G^3+140 M_p^4 \eta ^2 G^6\right)}{ a^4 S^2_{BH}}$ \\  \\
$
+\frac{ln \left(\frac{\sqrt{S_{BH}}M_{p}}{2\sqrt{\pi}}\right) \left(4 \pi  M_p^2 \eta  G^2\right) \left(a^6+6 a^4 M_p^2 \eta  G^3+40 a^2 M_p^4 \eta ^2 G^6+280 M_p^6 \eta ^3 G^9\right)}{a^6} $ \\ \\
$
-\frac{\pi  \left(a^8+4 a^6 M_p^2 \eta  G^3+24 a^4 M_p^4 \eta ^2 G^6+160 a^2 M_p^6 \eta ^3 G^9+1120 M_p^8 \eta ^4 G^{12}\right) ln \left(\frac{a^2\pi -2 G^3 M_{p}^2S_{BH}}{\pi}\right)}{2 a^6 G} $ \\ \\
$
= -\frac{70 \pi^4  M_p^2 \eta ^4 G^2}{3\left(\frac{A}{4}\right)^3} -\frac{10 \pi^3  M_p^2 \eta^3 G^2 \left(a^2+7 M_p^2 \eta  G^3\right)}{ a^2 \left(\frac{A}{4}\right)^2}- \frac{2\pi^2  M_p^2 \eta^2 G^2 \left(3 a^4+20 a^2 M_p^2 \eta  G^3+140 M_p^4 \eta ^2 G^6\right)}{ a^4 \left(\frac{A}{4}\right)^2}$ \\ \\
$
+\frac{ln \left(\frac{\sqrt{A}M_{p}}{4\sqrt{\pi}}\right) \left(4 \pi  M_p^2 \eta  G^2\right) \left(a^6+6 a^4 M_p^2 \eta  G^3+40 a^2 M_p^4 \eta ^2 G^6+280 M_p^6 \eta ^3 G^9\right)}{a^6} $ \\
\begin{equation}
-\frac{\pi  \left(a^8+4 a^6 M_p^2 \eta  G^3+24 a^4 M_p^4 \eta ^2 G^6+160 a^2 M_p^6 \eta ^3 G^9+1120 M_p^8 \eta ^4 G^{12}\right) ln \left(\frac{a^2\pi - G^3 M_{p}^2\left(\frac{A}{2}\right)}{\pi}\right)}{2 a^6 G} 
\end{equation}
\begin{figure}[h]
\begin{center}
Fig.10.1~~~~~~~~~~~~\hspace{4cm}~~~~~~~~Fig.10.2\\
\includegraphics[height=2.5in, width=3.2in]{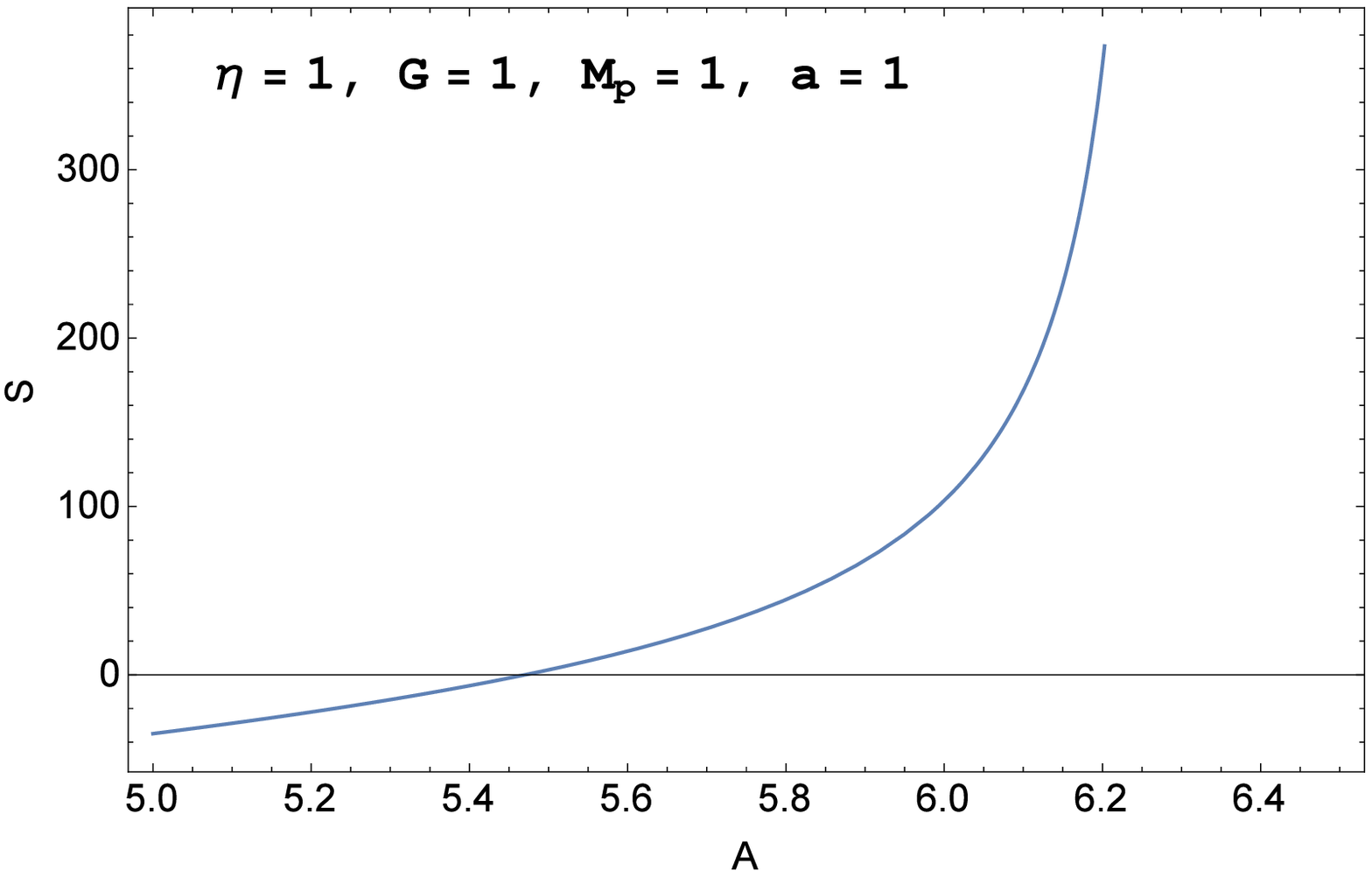}~~~~
\includegraphics[height=2.5in, width=3.2in]{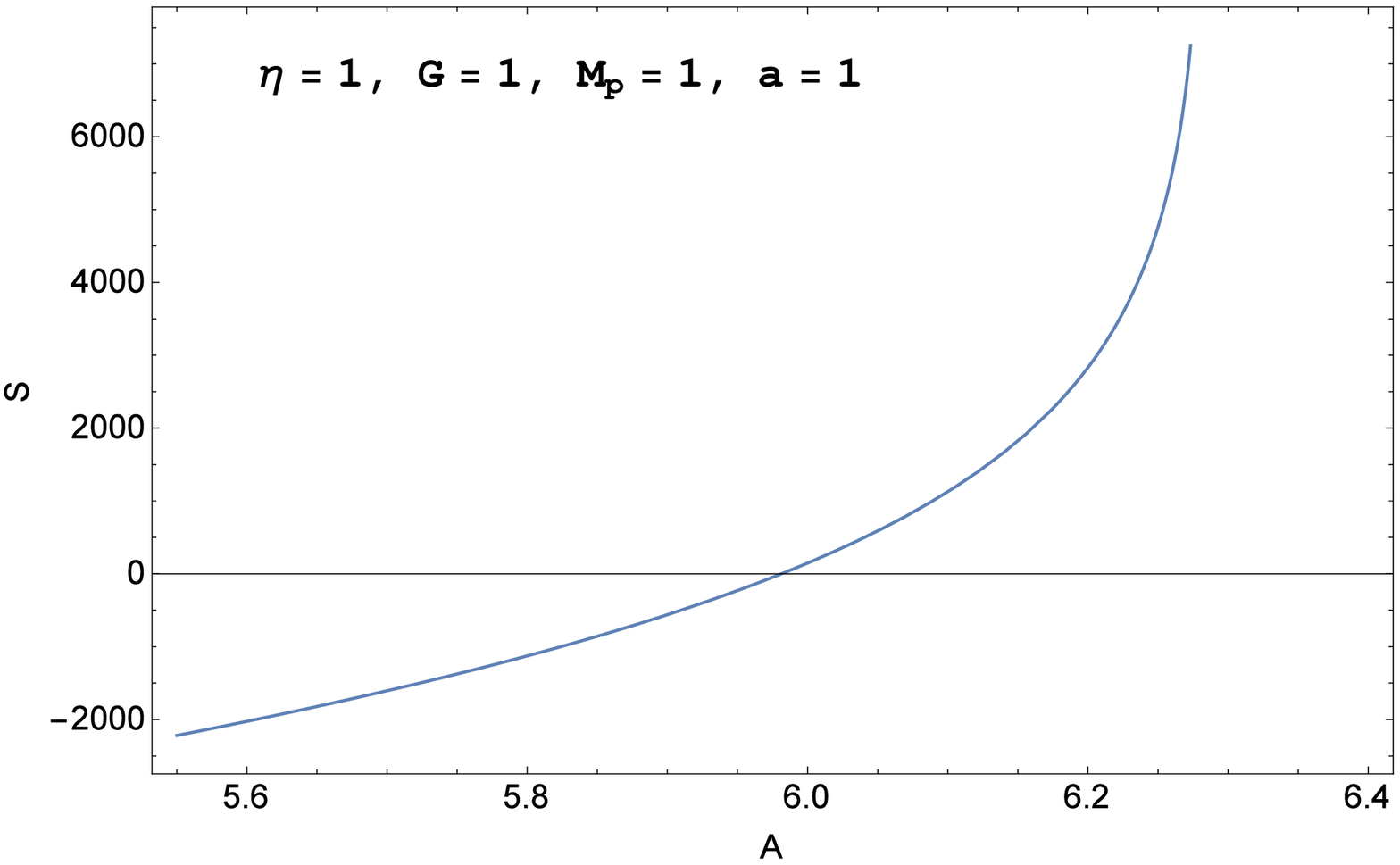}\\
\end{center}
Fig.10.1 and 10.2 shows that for $n=1$ the BH starts its journey faster than for the $n=2$ case. Initially for $n=1$ $S$ increases slowly with increase in $A$ but then the rate of increase keeps increasing. In both the cases, $S$ increases as $A$ increase.
\end{figure}
\pagebreak \\
$
{\large \textbf{Case-2} : w_{q}=-\frac{2}{3}}$ \\ \\
$ S=8\pi G^2 a^2 r_{q}\int \frac{M +\frac{\eta  M^n_p}{2M^{n-1}}+\frac{3 \eta ^2 M_p^{2n}}{8 M^{2n-1}}+\frac{5 \eta ^3 M_p^{3n}}{16 M^{3n-1}}+\frac{35 \eta ^4 M_p^{4n}}{128 M^{4n-1}}}{a^2 r_q-8 G^3 M^2 r_q-2 a^2 G^2 M}$ \\ \\ 
{\large\textbf{Case-2(a)}} : n=1 \\ \\
\resizebox{.97 \textwidth}{!} 
{
$
S=-\frac{35 \pi  \eta ^4 G^2 M_p^4}{32 M^2}-\frac{\pi  G^2 \left(140 a^2 \eta ^4 G^2 M_p^4 r_q+80 a^2 \eta ^3 M_p^3 r_q^2\right)}{32 a^2 r_q^2 M}+\frac{\pi  \eta ^2 G^2 M_p^2 \log (M) \left(a^2 \left(35 \eta ^2 G^4 M_p^2+20 \eta  G^2 M_p r_q+12 r_q^2\right)+70 \eta ^2 G^3 M_p^2 r_q^2\right)}{4 a^2 r_q^2}$
} \\ \\
$-\frac{\pi  \log \left(a^2 \left(2 G^2 M-r_q\right)+8 G^3 M^2 r_q\right) \left(4 a^4 r_q^2+a^2 \eta ^2 G^3 M_p^2 \left(35 \eta ^2 G^4 M_p^2+20 \eta  G^2 M_p r_q+12 r_q^2\right)+70 \eta ^4 G^6 M_p^4 r_q^2\right)}{8 a^2 r_q^2 G}+$ \\ \\
\resizebox{.97 \textwidth}{!} 
{
$\frac{\pi  \tanh ^{-1}\left(\frac{\sqrt{G} \left(a^2+8 G M r_q\right)}{a \sqrt{a^2 G+8 r_q^2}}\right) \left(-4 a^5 r_q^2+a^3 \eta  G M_p \left(35 \eta ^3 G^6 M_p^3+20 \eta ^2 G^4 M_p^2 r_q+12 \eta  G^2 M_p r_q^2+16 r_q^3\right)+10 a \eta ^3 G^4 M_p^3 r_q^2 \left(21 \eta  G^2 M_p+8 r_q\right)\right)}{4 a^2 \sqrt{G} r_q^2 \sqrt{a^2 G+8 r_q^2}}
$ 
}\\ \\
\resizebox{.97 \textwidth}{!} 
{
$
=-\frac{35 \pi^2  \eta ^4 G^2 M_p^2}{8 S_{BH}}-\frac{5\pi^{\frac{3}{2}}\eta^3 G^2 M^2_{p}\left[ 7\eta G^2 M_{p} +4 r_{q} \right]}{4 r_{q} \sqrt{S_{BH}}} +\frac{\pi  \eta ^2 G^2 M_p^2 \log \left(\frac{\sqrt{S_{BH}}M_{p}}{2\sqrt{\pi}}\right) \left(a^2 \left(35 \eta ^2 G^4 M_p^2+20 \eta  G^2 M_p r_q+12 r_q^2\right)+70 \eta ^2 G^3 M_p^2 r_q^2\right)}{4 a^2 r_q^2}$ 
} \\ \\
$-\frac{\pi  \log  \left(2a^2 G^2 \left(\frac{\sqrt{S_{BH}}M_{p}}{2\sqrt{\pi}}\right)-a^2r_q +8 G^3 \left(\frac{S_{BH}M_{p}}{4\pi}\right) r_q\right) \left(4 a^4 r_q^2+a^2 \eta ^2 G^3 M_p^2 \left(35 \eta ^2 G^4 M_p^2+20 \eta  G^2 M_p r_q+12 r_q^2\right)+70 \eta ^4 G^6 M_p^4 r_q^2\right)}{8 a^2 r_q^2 G}$ \\ \\
\resizebox{.97 \textwidth}{!} 
{
$+ \frac{\pi  \tanh ^{-1}\left(\frac{ a^2 \sqrt{G}+8 G^{\frac{3}{2}} r_q \left(\frac{\sqrt{S_{BH}}M_{p}}{2\sqrt{\pi}} \right)}{a \sqrt{a^2 G+8 r_q^2}}\right) \left[-4 a^5 r_q^2+a^3 \eta  G M_p \left(35 \eta ^3 G^6 M_p^3+20 \eta ^2 G^4 M_p^2 r_q+12 \eta  G^2 M_p r_q^2+16 r_q^3\right)+10 a \eta ^3 G^4 M_p^3 r_q^2 \left(21 \eta  G^2 M_p+8 r_q\right)\right]}{4 a^2 \sqrt{G} r_q^2 \sqrt{a^2 G+8 r_q^2}}
$
} \\ \\
\resizebox{.97 \textwidth}{!} 
{
$
=-\frac{35 \pi^2  \eta ^4 G^2 M_p^2}{8 \left(\frac{A}{4}\right)}-\frac{5\pi^{\frac{3}{2}}\eta^3 G^2 M^2_{p}\left[ 7\eta G^2 M_{p} +4 r_{q} \right]}{4 r_{q} \sqrt{\left(\frac{A}{4}\right)}} +\frac{\pi  \eta ^2 G^2 M_p^2 \log \left(\frac{\sqrt{\left(\frac{A}{4}\right)}M_{p}}{2\sqrt{\pi}}\right) \left(a^2 \left(35 \eta ^2 G^4 M_p^2+20 \eta  G^2 M_p r_q+12 r_q^2\right)+70 \eta ^2 G^3 M_p^2 r_q^2\right)}{4 a^2 r_q^2}$ 
} \\ \\
$-\frac{\pi  \log  \left(2a^2 G^2 \left(\frac{\sqrt{\left(\frac{A}{4}\right)}M_{p}}{2\sqrt{\pi}}\right)-a^2r_q +8 G^3 \left(\frac{\left(\frac{A}{4}\right) M_{p}}{4\pi}\right) r_q\right) \left(4 a^4 r_q^2+a^2 \eta ^2 G^3 M_p^2 \left(35 \eta ^2 G^4 M_p^2+20 \eta  G^2 M_p r_q+12 r_q^2\right)+70 \eta ^4 G^6 M_p^4 r_q^2\right)}{8 a^2 r_q^2 G}$ \\ \\
\resizebox{.97 \textwidth}{!} 
{
$
+ \pi  \tanh ^{-1}\left(\frac{ a^2 \sqrt{G}+8 G^{\frac{3}{2}} r_q \left(\frac{\sqrt{\left(\frac{A}{4}\right)}M_{p}}{2\sqrt{\pi}} \right)}{a \sqrt{a^2 G+8 r_q^2}}\right) \left[-4 a^5 r_q^2+a^3 \eta  G M_p \left(35 \eta ^3 G^6 M_p^3+20 \eta ^2 G^4 M_p^2 r_q+12 \eta  G^2 M_p r_q^2+16 r_q^3\right)\right. $
}\\
\begin{equation}
\resizebox{.57 \textwidth}{!} 
{
$\left.+10 a \eta ^3 G^4 M_p^3 r_q^2 \left(21 \eta  G^2 M_p+8 r_q\right)\right]\left[4 a^2 \sqrt{G} r_q^2 \sqrt{a^2 G+8 r_q^2}\right]^{-1}$
}
\end{equation}\\

{\large\textbf{Case-2(b)}} : n=2 \\ \\
Considering terms upto $O(\eta^2)$ we get :\\ \\
$ S= -\frac{3 \pi  \eta ^2 G^2 M_p^4}{2 M^2} -\frac{6 \pi  \eta ^2 G^4 M_p^4}{M r_q} -\frac{\pi  \tanh ^{-1}\left(\frac{\sqrt{G} \left(a^2+8 G M r_q\right)}{a \sqrt{a^2 G+8 r_q^2}}\right) \left(a^4 r_q^2-4 a^2 \eta  G^3 M_p^2 \left(3 \eta  G^4 M_p^2+r_q^2\right)-72 \eta ^2 G^6 M_p^4 r_q^2\right)}{a \sqrt{G} r_q^2 \sqrt{a^2 G+8 r_q^2}}$ \\ \\
\resizebox{.97 \textwidth}{!} 
{
$
+\frac{\log (M) \left(4 \pi  \eta  G^2 M_p^2\right) \left(3 a^2 \eta  G^4 M_p^2+r_q^2 \left(a^2+6 \eta  G^3 M_p^2\right)\right)}{a^2 r_q^2} -\frac{\pi  \log \left(a^2 \left(2 G^2 M-r_q\right)+8 G^3 M^2 r_q\right) \left(a^4 r_q^2+4 a^2 \eta  G^3 M_p^2 \left(3 \eta  G^4 M_p^2+r_q^2\right)+24 \eta ^2 G^6 M_p^4 r_q^2\right)}{2 a^2 G r_q^2}
$
} \\ \\
\resizebox{.97 \textwidth}{!} 
{
$
= -\frac{6 \pi^2  \eta ^2 G^2 M_p^2}{S_{BH}} -\frac{12 \pi^{\frac{3}{2}}  \eta ^2 G^4 M_p^3}{\sqrt{S_{BH}} r_q} -\frac{\pi  \tanh ^{-1}\left(\frac{\sqrt{G} \left(a^2+4 G \left(\frac{M_{p}\sqrt{S_{BH}}}{\sqrt{\pi}}\right) r_q\right)}{a \sqrt{a^2 G+8 r_q^2}}\right) \left(a^4 r_q^2-4 a^2 \eta  G^3 M_p^2 \left(3 \eta  G^4 M_p^2+r_q^2\right)-72 \eta ^2 G^6 M_p^4 r_q^2\right)}{a \sqrt{G} r_q^2 \sqrt{a^2 G+8 r_q^2}}$
} \\ \\
\resizebox{.97 \textwidth}{!} 
{
$
+\frac{\log \left( \frac{M_{p}\sqrt{S_{BH}}}{2\sqrt{\pi}} \right) \left(4 \pi  \eta  G^2 M_p^2\right) \left(3 a^2 \eta  G^4 M_p^2+r_q^2 \left(a^2+6 \eta  G^3 M_p^2\right)\right)}{a^2 r_q^2}
-\frac{\pi  \log \left(a^2 \left( \frac{G^2 M_{p}\sqrt{S_{BH}}}{\sqrt{\pi}} -r_q\right)+ \frac{2 G^3 M^2_{p} r_q S_{BH}}{\pi}\right) \left(a^4 r_q^2+4 a^2 \eta  G^3 M_p^2 \left(3 \eta  G^4 M_p^2+r_q^2\right)+24 \eta ^2 G^6 M_p^4 r_q^2\right)}{2 a^2 G r_q^2}
$
}\\ \\
\resizebox{.97 \textwidth}{!} 
{
$
= -\frac{6 \pi^2  \eta ^2 G^2 M_p^2}{\left(\frac{A}{4}\right)} -\frac{12 \pi^{\frac{3}{2}}  \eta ^2 G^4 M_p^3}{\sqrt{\left(\frac{A}{4}\right)} r_q} -\frac{\pi  \tanh ^{-1}\left(\frac{\sqrt{G} \left(a^2+ \frac{4 G r_q M_{p}\sqrt{\left(\frac{A}{4}\right)}}{\sqrt{\pi}} \right)}{a \sqrt{a^2 G+8 r_q^2}}\right) \left(a^4 r_q^2-4 a^2 \eta  G^3 M_p^2 \left(3 \eta  G^4 M_p^2+r_q^2\right)-72 \eta ^2 G^6 M_p^4 r_q^2\right)}{a \sqrt{G} r_q^2 \sqrt{a^2 G+8 r_q^2}}
$
}\\ \\
\begin{equation}
\resizebox{.97 \textwidth}{!} 
{
$
+\frac{\log \left( \frac{M_{p}\sqrt{\left(\frac{A}{4}\right)}}{2\sqrt{\pi}} \right) \left(4 \pi  \eta  G^2 M_p^2\right) \left(3 a^2 \eta  G^4 M_p^2+r_q^2 \left(a^2+6 \eta  G^3 M_p^2\right)\right)}{a^2 r_q^2}
 -\frac{\pi  \log \left(a^2 \left( \frac{G^2 M_{p}\sqrt{\left(\frac{A}{4}\right)}}{\sqrt{\pi}} -r_q\right)+ \frac{2 G^3 M^2_{p} r_q \left(\frac{A}{4}\right)}{\pi}\right) \left(a^4 r_q^2+4 a^2 \eta  G^3 M_p^2 \left(3 \eta  G^4 M_p^2+r_q^2\right)+24 \eta ^2 G^6 M_p^4 r_q^2\right)}{2 a^2 G r_q^2} $
 }
\end{equation} 
\begin{figure}[h]
\begin{center}
Fig.11.1~~~~~~~~~~\hspace{4cm}~~~~~~~~~~~ Fig.11.2\\
\includegraphics[height=2.5in, width=3.2in]{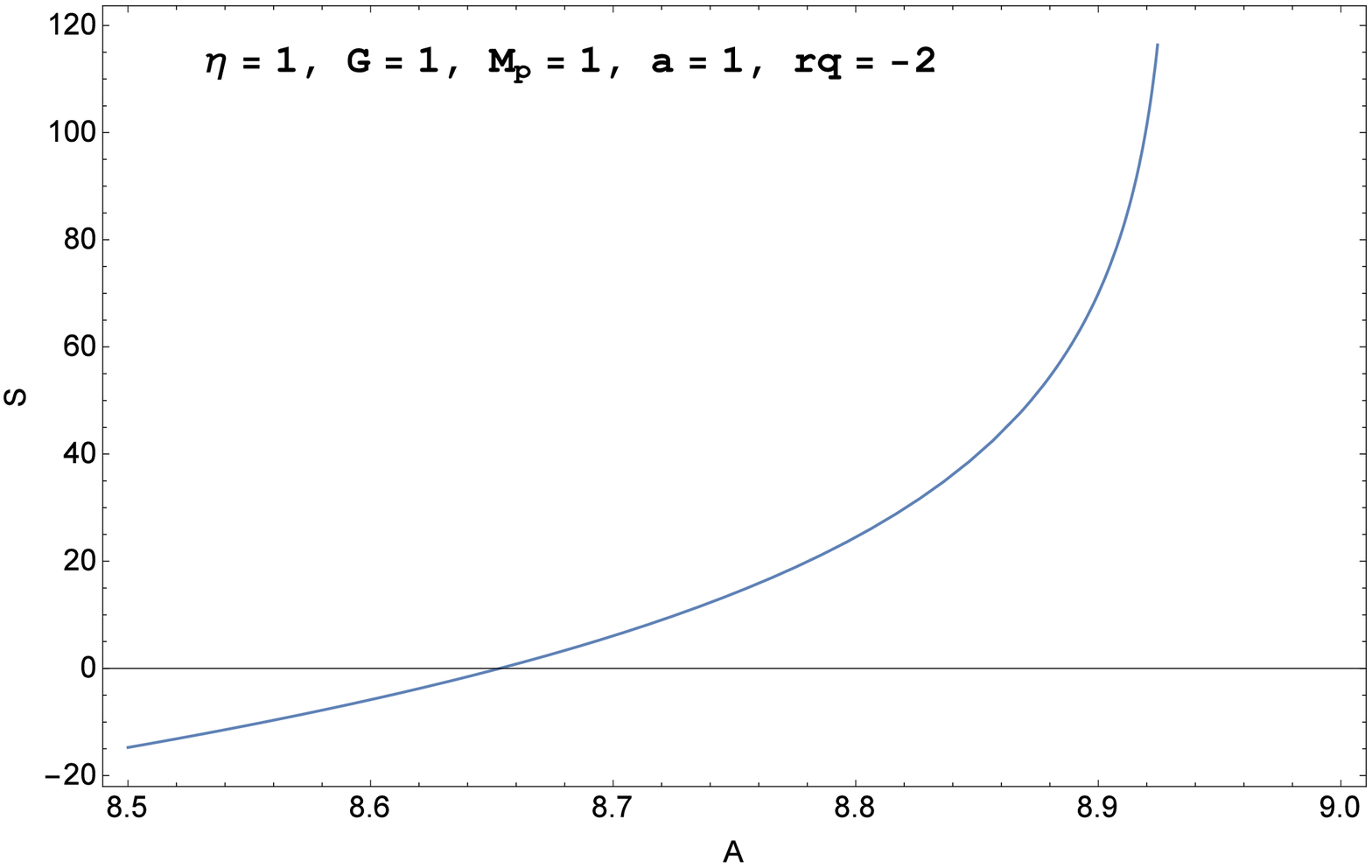}~~~~
\includegraphics[height=2.5in, width=3.2in]{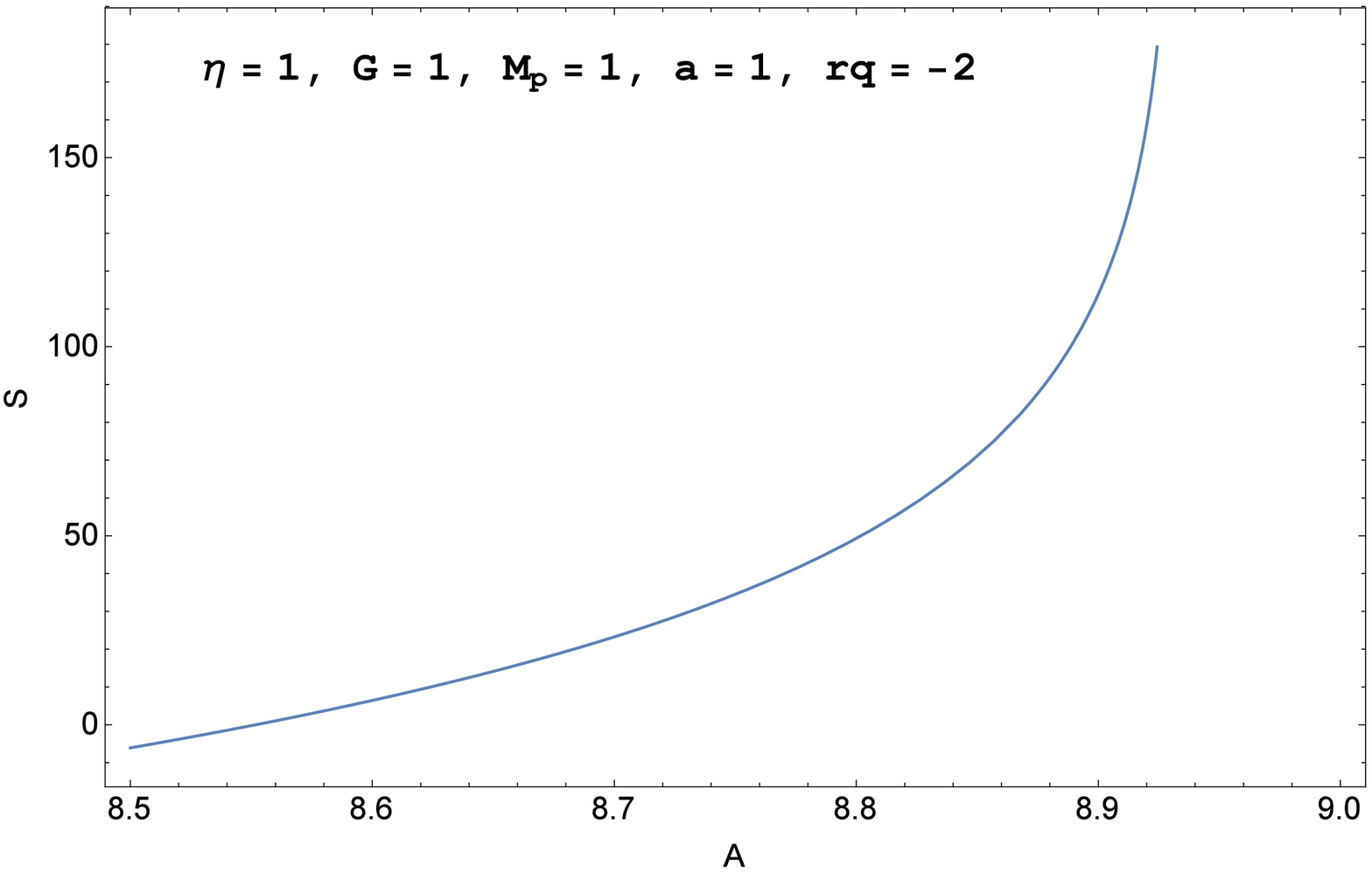}\\
\end{center}
Fig.11.1 and 11.2 depicts the $S$ vs $A$ curves for $n=1,2$ respectively. For $n=1$ case the BH starts its journey late than the $n=2$ case. In both the cases the $S$ vs $A$ curve is almost similar. We get a feasible graph only for values of $r_q < 0$. 
\end{figure}
\pagebreak \\
$
{\large \textbf{Case-3} : w_{q}= -1}$ \\ \\
$ S=\mathlarger{ 8\pi G^2 a^2 r^2_{q}\int \frac{M +\frac{\eta  M^n_p}{2M^{n-1}}+\frac{3 \eta ^2 M_p^{2n}}{8 M^{2n-1}}+\frac{5 \eta ^3 M_p^{3n}}{16 M^{3n-1}}+\frac{35 \eta ^4 M_p^{4n}}{128 M^{4n-1}}}{a^2 r^2_q-8M^2 G^3 r^2_q -8 M^2 G^3 a^2 }dM}$\\ \\
{\large\textbf{Case-3(a)}} : n=1 \\ \\
\resizebox{.97 \textwidth}{!} 
{$
S= -\frac{35 \pi  \eta ^4 G^2 M_p^4}{32 M^2} -\frac{5 \pi  \eta ^3 G^2 M_p^3}{2 M} +\frac{\sqrt{2} \pi  \eta  \sqrt{G} M_p \tanh ^{-1}\left(\frac{2 \sqrt{2} G^{3/2} M \sqrt{a^2+r_q^2}}{a r_q}\right) \left(a^2 \left(5 \eta ^2 G^3 M_p^2+r_q^2\right)+5 \eta ^2 G^3 M_p^2 r_q^2\right)}{a r_q \sqrt{a^2+r_q^2}}
$
}\\ \\
\resizebox{.97 \textwidth}{!} 
{
$
-\frac{\pi  \log \left[a^2 \left(r_q^2-8 G^3 M^2\right)-8 G^3 M^2 r_q^2\right] \left(a^4 \left(35 \eta ^4 G^6 M_p^4+6 \eta ^2 G^3 M_p^2 r_q^2+2 r_q^4\right)+a^2 \left(70 \eta ^4 G^6 M_p^4 r_q^2+6 \eta ^2 G^3 M_p^2 r_q^4\right)+35 \eta ^4 G^6 M_p^4 r_q^4\right)}{4 a^2 G r_q^2 \left(a^2+r_q^2\right)}
$
}\\ \\
\resizebox{.57 \textwidth}{!} 
{
$
+\frac{\pi  G^2 \log (M) \left(a^2 \left(35 \eta ^4 G^3 M_p^4+6 \eta ^2 M_p^2 r_q^2\right)+35 \eta ^4 G^3 M_p^4 r_q^2\right)}{2 a^2 r_q^2}
$
}\\ \\
\resizebox{.97 \textwidth}{!} 
{
$
= -\frac{35 \pi^2  \eta ^4 G^2 M_p^2}{8 S_{BH}} -\frac{5 \pi^{\frac{3}{2}}  \eta ^3 G^2 M_p^2}{\sqrt{S_{BH}}} +\frac{\sqrt{2} \pi  \eta  \sqrt{G} M_p \tanh ^{-1}\left(\frac{ \sqrt{2} G^{3/2} \sqrt{S_{BH}}M_{p} \sqrt{a^2+r_q^2}}{\pi a r_q}\right) \left(a^2 \left(5 \eta ^2 G^3 M_p^2+r_q^2\right)+5 \eta ^2 G^3 M_p^2 r_q^2\right)}{a r_q \sqrt{a^2+r_q^2}}
$
}\\ \\
\resizebox{.97 \textwidth}{!} 
{
$
-\frac{\pi  \log \left[a^2 \left(r_q^2- \frac{2G^3 M^2_{p}S_{BH}}{\pi}\right)-\frac{2G^3 M^2_{p}r^2_{q}S_{BH}}{\pi}\right] \left(a^4 \left(35 \eta ^4 G^6 M_p^4+6 \eta ^2 G^3 M_p^2 r_q^2+2 r_q^4\right)+a^2 \left(70 \eta ^4 G^6 M_p^4 r_q^2+6 \eta ^2 G^3 M_p^2 r_q^4\right)+35 \eta ^4 G^6 M_p^4 r_q^4\right)}{4 a^2 G r_q^2 \left(a^2+r_q^2\right)}
$
}\\ \\
\resizebox{.57 \textwidth}{!} 
{
$
+\frac{\pi  G^2 \log \left( \frac{ M_{p}\sqrt{S_{BH}}}{2\sqrt{\pi}} \right) \left(a^2 \left(35 \eta ^4 G^3 M_p^4+6 \eta ^2 M_p^2 r_q^2\right)+35 \eta ^4 G^3 M_p^4 r_q^2\right)}{2 a^2 r_q^2}
$
}\\ \\
\resizebox{.97 \textwidth}{!} 
{
$
= -\frac{35 \pi^2  \eta ^4 G^2 M_p^2}{8 \left(\frac{A}{4}\right)} -\frac{5 \pi^{\frac{3}{2}}  \eta ^3 G^2 M_p^2}{\sqrt{\left(\frac{A}{4}\right)}} +\frac{\sqrt{2} \pi  \eta  \sqrt{G} M_p \tanh ^{-1}\left(\frac{ \sqrt{2} G^{3/2} \sqrt{\left(\frac{A}{4}\right)}M_{p} \sqrt{a^2+r_q^2}}{\pi a r_q}\right) \left(a^2 \left(5 \eta ^2 G^3 M_p^2+r_q^2\right)+5 \eta ^2 G^3 M_p^2 r_q^2\right)}{a r_q \sqrt{a^2+r_q^2}}
$
}\\ \\
\resizebox{.97 \textwidth}{!} 
{
$
-\frac{\pi  \log \left[a^2 \left(r_q^2- \frac{2G^3 M^2_{p}\left(\frac{A}{4}\right)}{\pi}\right)-\frac{2G^3 M^2_{p}r^2_{q}\left(\frac{A}{4}\right)}{\pi}\right] \left(a^4 \left(35 \eta ^4 G^6 M_p^4+6 \eta ^2 G^3 M_p^2 r_q^2+2 r_q^4\right)+a^2 \left(70 \eta ^4 G^6 M_p^4 r_q^2+6 \eta ^2 G^3 M_p^2 r_q^4\right)+35 \eta ^4 G^6 M_p^4 r_q^4\right)}{4 a^2 G r_q^2 \left(a^2+r_q^2\right)}
$
}
\begin{equation}
\resizebox{.62 \textwidth}{!} 
{
$
+\frac{\pi  G^2 \log \left( \frac{ M_{p}\sqrt{\left(\frac{A}{4}\right)}}{2\sqrt{\pi}} \right) \left(a^2 \left(35 \eta ^4 G^3 M_p^4+6 \eta ^2 M_p^2 r_q^2\right)+35 \eta ^4 G^3 M_p^4 r_q^2\right)}{2 a^2 r_q^2}
$
}
\end{equation}\\

{\large\textbf{Case-3(b)}} : n=2 \\
Considering terms upto $O(\eta^2)$ we get :\\ \\
\resizebox{.7 \textwidth}{!} 
{$
S= -\frac{3 \pi  \eta ^2 G^2 M_p^4}{2 M^2}  +\frac{4 \pi  \eta  G^2 M_p^2}{a^2 r_q^2}\log (M) \left[a^2 \left(6 \eta  G^3 M_p^2+r_q^2\right)+6 \eta  G^3 M_p^2 r_q^2\right]$
}\\ \\
$
-\frac{\pi  \log \left[a^2 \left(r_q^2-8 G^3 M^2\right)-8 G^3 M^2 r_q^2\right] \left[24 a^4 \eta ^2 G^6 M_p^4+4 a^2 \eta  G^3 M_p^2 r_q^2 \left(a^2+12 \eta  G^3 M_p^2\right)+r_q^4 \left(a^4+4 a^2 \eta  G^3 M_p^2+24 \eta ^2 G^6 M_p^4\right)\right]}{2 a^2 G r_q^2 \left(a^2+r_q^2\right)}
$\\ \\
\resizebox{.77 \textwidth}{!} 
{
$
=-\frac{6\pi^2 \eta^2 G^2 M^2_{p}}{S_{BH}}+ \frac{4 \pi  \eta  G^2 M_p^2}{a^2 r_q^2}\log \left( \frac{M_p \sqrt{S_{BH}}}{2\sqrt{\pi}} \right) \left[a^2 \left(6 \eta  G^3 M_p^2+r_q^2\right)+6 \eta  G^3 M_p^2 r_q^2\right]$
}\\ \\
\resizebox{.97 \textwidth}{!} 
{
$
-\frac{\pi  \log \left[a^2 \left(r_q^2- \frac{2G^3 M^2_{p}S_{BH}}{\pi}\right)-\left( \frac{2G^3 M^2_{p}r^2_{q}S_{BH}}{\pi} \right)\right] \left[24 a^4 \eta ^2 G^6 M_p^4+4 a^2 \eta  G^3 M_p^2 r_q^2 \left(a^2+12 \eta  G^3 M_p^2\right)+r_q^4 \left(a^4+4 a^2 \eta  G^3 M_p^2+24 \eta ^2 G^6 M_p^4\right)\right]}{2 a^2 G r_q^2 \left(a^2+r_q^2\right)}
$
}\\ \\
\resizebox{.8 \textwidth}{!} 
{
$
=-\frac{6\pi^2 \eta^2 G^2 M^2_{p}}{\left(\frac{A}{4}\right)}+ \frac{4 \pi  \eta  G^2 M_p^2}{a^2 r_q^2}\log \left( \frac{M_p \sqrt{\left(\frac{A}{4}\right)}}{2\sqrt{\pi}} \right) \left[a^2 \left(6 \eta  G^3 M_p^2+r_q^2\right)+6 \eta  G^3 M_p^2 r_q^2\right]$ 
}\\
\resizebox{.84 \textwidth}{!} 
{
$
-\pi  \log \left[a^2 \left(r_q^2- \frac{2G^3 M^2_{p}\left(\frac{A}{4}\right)}{\pi}\right)-\left( \frac{2G^3 M^2_{p}r^2_{q}\left(\frac{A}{4}\right)}{\pi} \right)\right] \left[24 a^4 \eta ^2 G^6 M_p^4+4 a^2 \eta  G^3 M_p^2 r_q^2 \left(a^2+12 \eta  G^3 M_p^2\right)\right. $
}\\
\begin{equation}
\resizebox{.6 \textwidth}{!} 
{
$
\left.+r_q^4 \left(a^4+4 a^2 \eta  G^3 M_p^2+24 \eta ^2 G^6 M_p^4\right)\right]\left[2 a^2 G r_q^2 \left(a^2+r_q^2\right)\right]^{-1}$
}
\end{equation}
\begin{figure}[h]
\begin{center}
Fig.12.1~~~~~~~~~~~\hspace{4cm}~~~~~~~~~~~~ Fig.12.2\\
\includegraphics[height=2.5in, width=3.2in]{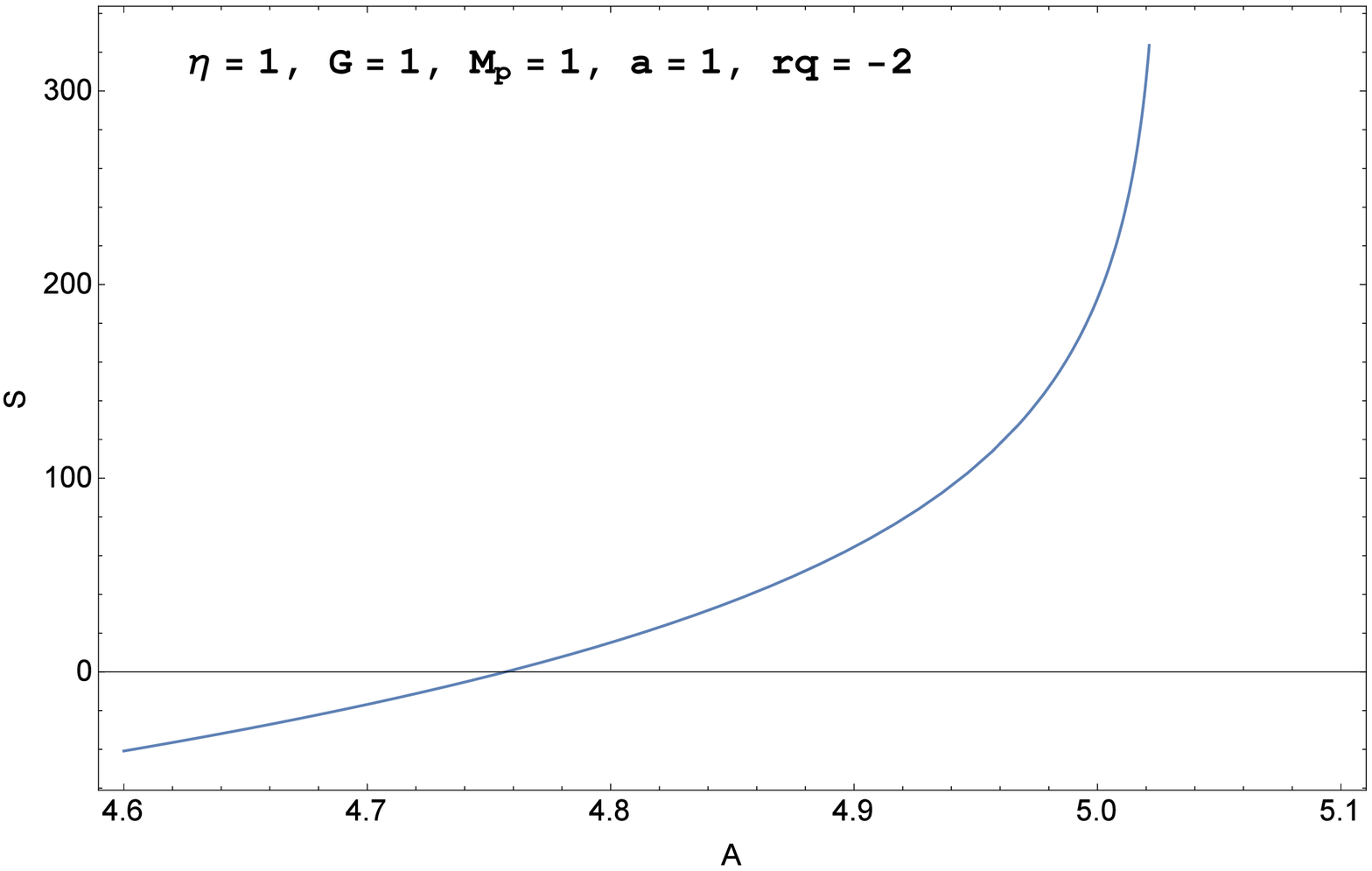}~~~~
\includegraphics[height=2.5in, width=3.2in]{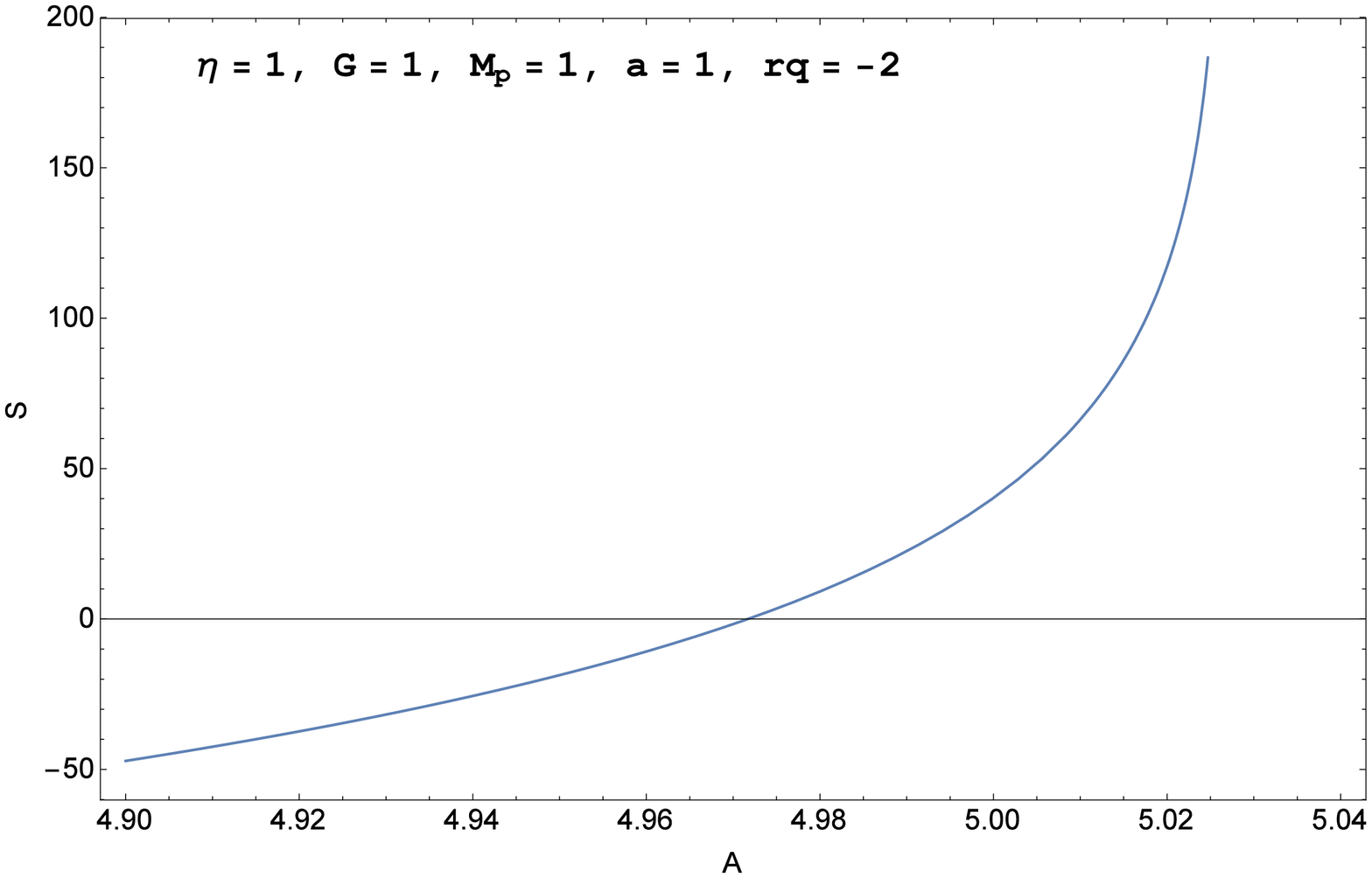}\\
\end{center}
Fig.12.1 and 12.2 depicts the $S$ vs $A$ curve for $n=1$ and $n=2$ respectively. The graphs are very similar to Fig.10.1 and 10.2. For the $n=1$ case the BH starts its journey before than the $n=2$ case. We get a feasible graph only for values of $r_q < 0$. 
\end{figure}
\pagebreak \\
\section{Conclusion}
In this paper we studied the thermodynamic effects on a black hole under the influence of rainbow gravity. Firstly we considered the rainbow surface gravity and derived the rainbow Hawking temperature, and subsequently we obtained the other thermodynamic quantities such as heat capacity and entropy. In the process we came across the existence of remnant mass and critical mass and also derived the critical points of the black hole thermodynamic ensemble and studied its stability. \\
We observed that the temperature of the black hole in rainbow gravity depends on the energy E of a probe provided by the rainbow functions. We derived the remnant mass of the black hole and showed that the Hawking temperature becomes zero at Planck scale. This implies that the divergent nature of the standard Hawking temperature is regularized by the rainbow gravity and it also prevents the complete evaporation of the black hole which can be treated as a probable candidate for solving the information loss paradox and naked singularity problems of black holes.\\
We derive the entropy of the black holes using the rainbow Hawking temperature and thus the entropy expressions consists of quantum corrections. Then the entropy relation is expressed in terms of the area of the event horizon of the black hole and from the area-entropy  graphs we make several meaningful interpretations and closely  analyse the effects of the different values of the rainbow parameters on these thermodynamic properties.\\

It can be stated from entropy vs. area of horizon graphs that the presence of the rainbow parameter $\eta$ in the rainbow function affects the temperature of the system as increased values of $\eta$ decreases the temperature. Hence it can be said that the rainbow parameter $\eta$ is somehow a representative of less temperature.
In case of Reissner Nordstrom de Sitter black hole surrounded by quintessence for $w_q = -\frac{2}{3}$ and $-1$ it can be observed that, for non-negative values of $r_q$ there can be no feasible graphical representations of S vs. A graphs. Hence for physically meaningful interpretations the quantity $\left(\frac{r_q}{r}\right)^{3w_q+1}$ must be a negative quantity  in solution of the RNDS BH surrounded by the quintessence.
It is interesting to observe that the introduction of rainbow gravity keeps the overall physical interpretations similar to the Einstein gravity but it gives a quantum correction to the latter at the Planck scale and gives us new and interesting insights into the thermodynamic journey of a BH.  
\newpage
\addcontentsline{toc}{part}{\bf  Bibliography}
 \baselineskip
.81cm \markright{ Bibliography}


\begin{thebibliography}{450}
\bibitem{1} G. Amelino-Camelia, “Testable scenario for relativity with minimum-length,” Phys. Lett. B 510 , 255 (2001) [arXiv:hep-th/0012238].
\bibitem{2} G. Amelino-Camelia, “Relativity in space-times with short-distance structure governed by an observer-independent (Planckian) length scale,” Int. J. Mod. Phys.D11, 35 (2002) [arXiv:gr-qc/0012051].
\bibitem{3} J. Kowalski-Glikman, “Observer independent quantum of mass,” Phys. Lett. A286, 391 (2001) [arXiv:hep-th/0102098].
\bibitem{4} N. R. Bruno, G. Amelino-Camelia and J. Kowalski-Glikman, “De-formed boost transformations that saturate at the Planck scale,” Phys. Lett. B522, 133 (2001) [arXiv:hep-th/0107039].
\bibitem{5} J. Magueijo and L. Smolin, “Generalized Lorentz invariance with an
invariant energy scale,” Phys. Rev. D67, 044017 (2003) [gr-qc/0207085].
\bibitem{6} J. Magueijo and L. Smolin, “Gravity’s rainbow,” Class. Quant. Grav. 21, 1725 (2004) [gr-qc 0305055].
\bibitem{7} S. Liberati, S. Sonego and M. Visser, “Interpreting doubly special relativity as a modified theory of measurement,” Phys. Rev. D71, 045001(2005) [gr-qc/0410113].
\bibitem{8} P. Galan and G. A. Mena Marugan, “Quantum time uncertainty in a gravity’s rainbow formalism,” Phys. Rev. D70, 124003 (2004) [gr-qc/0411089].
\bibitem{9} P. Galan and G. A. Mena Marugan, “Length uncertainty in a gravity’s rainbow formalism,” Phys. Rev. D72, 044019 (2005) [gr-qc/0507098].
\bibitem{10} J. Hackett, “Asymptotic flatness in rainbow gravity,” Class. Quant. Grav.23, 3833 (2006) [gr-qc/0509103].
\bibitem{11} Y. Ling, “Rainbow universe,” JCAP 0708, 017 (2007) [gr-qc/0609129].
\bibitem{12} Y. Ling, S. He and H. -b. Zhang, “The Kinematics of particles moving in rainbow spacetime,” Mod. Phys. Lett. A22, 2931 (2007) [gr-qc/0609130].
\bibitem{13} F. Girelli, S. Liberati and L. Sindoni, “Planck-scale modified dispersion relations and Finsler geometry,” Phys. Rev. D75, 064015 (2007) [gr-qc/0611024].
\bibitem{14} Y. Ling and Q. Wu, “The Big Bounce in Rainbow Universe,” Phys. Lett. B687, 103 (2010) [arXiv:0811.2615 [gr-qc]].
\bibitem{15} R. Garattini and G. Mandanici, “Particle propagation and effective space-time in Gravity’s Rainbow,” Phys. Rev. D85, 023507 (2012)
[arXiv:1109.6563 [gr-qc]].
\bibitem{16} R. Garattini and G. Mandanici, “Modified Dispersion Relations lead to a finite Zero Point Gravitational Energy,” Phys. Rev. D83, 084021 (2011) [arXiv:1102.3803 [gr-qc]].
\bibitem{17} R. Garattini and F. S. N. Lobo, “Self-sustained wormholes in modified dispersion relations,” Phys. Rev. D85, 024043 (2012) [arXiv:1111.5729[gr-qc]].
\bibitem{18} R. Garattini, “Distorting General Relativity: Gravity’s Rainbow and f(R) theories at work,” JCAP1306, 017 (2013) [arXiv:1210.7760 [gr-qc]].
\bibitem{19} R. Garattini and M. Sakellariadou, “Does gravity’s rainbow induce inflation without an inflaton?,” Phys. Rev. D90, 043521 (2014) [arXiv:1212.4987 [gr-qc]].
\bibitem{20} B. Majumder, “Singularity Free Rainbow Universe,” Int. J. Mod. Phys. D22, 1342021 (2013) [arXiv:1305.3709 [gr-qc]].
\bibitem{21} G. Amelino-Camelia, M. Arzano, G. Gubitosi and J. Magueijo, “Rainbow gravity and scale-invariant fluctuations,” Phys. Rev. D88, 041303 (2013) [arXiv:1307.0745 [gr-qc]].
\bibitem{22} A. Awad, A. F. Ali and B. Majumder, “Nonsingular Rainbow Universes”, JCAP1310, 052 (2013) [arXiv:1308.4343 [gr-qc]].
\bibitem{23} J. D. Barrow and J. Magueijo, “Intermediate inflation from rainbow gravity,” Phys. Rev. D88, 103525 (2013) [arXiv:1310.2072 [astro-ph.CO]].
\bibitem{24} G. Santos, G. Gubitosi and G. Amelino-Camelia, “On the initial singularity problem in rainbow cosmology,” JCAP1508, 005 (2015) [arXiv:1502.02833 [gr-qc]].
\bibitem{25} G. G. Carvalho, I. P. Lobo and E. Bittencourt, “Extended disformal approach in the scenario of Rainbow Gravity,” Phys. Rev. D93, 044005 (2016) [arXiv:1511.00495 [gr-qc]].
\bibitem{26} A. Ashour, M. Faizal, A. F. Ali and F. Hammad, “Branes in Gravity’s Rainbow,” Eur. Phys. J. C 76, 264 (2016) [arXiv:1602.04926 [hep-th]].
\bibitem{27} A. F. Ali, M. Faizal and M. M. Khalil, “Absence of Black Holes at LHC due to Gravity’s Rainbow,” Phys. Lett. B743, 295 (2015) [arXiv:1410.4765 [hep-th]].
\bibitem{28} A. F. Ali, M. Faizal and M. M. Khalil, “Remnant for all Black Objects due to Gravity’s Rainbow,” Nucl. Phys. B894, 341 (2015) [arXiv:1410.5706[hep-th]].
\bibitem{29} Y. Gim and W. Kim, “Black Hole Complementarity in Gravity’s Rainbow,” JCAP 1505, 002 (2015) [arXiv:1501.04702 [gr-qc]].
\bibitem{30} S. H. Hendi and M. Faizal, “Black holes in Gauss-Bonnet gravitys rainbow,” Phys. Rev. D92, 044027 (2015) [arXiv:1506.08062 [gr-qc]].
\bibitem{31} B. Mu, P. Wang and H. Yang, “Thermodynamics and Luminosities of Rainbow Black Holes,” JCAP1511, 045 (2015) [arXiv:1507.03768 [gr-qc]].
\bibitem{32} S. H. Hendi, M. Faizal, B. E. Panah and S. Panahiyan, “Charged dila-tonic black holes in gravitys rainbow,” Eur. Phys. J. C 76, 296 (2016) [arXiv:1508.00234].
\bibitem{33} Y. Gim and W. Kim, “Hawking, fiducial, and free-fall temperature of black hole on gravity’s rainbow,” Eur. Phys. J. C 76, 166 (2016)[arXiv:1509.06846 [gr-qc]].
\bibitem{34} S. Gangopadhyay and A. Dutta, “Constraints on rainbow gravity functions from black hole thermodynamics,” arXiv:1606.08295 [gr-qc].
\bibitem{35}H. Li, Y. Ling and X. Han, “Modified (A)dS Schwarzschild black holes in Rainbow spacetime,” Class. Quant. Grav. 26, 065004 (2009) [arXiv:0809.4819 [gr-qc]].
\bibitem{36}Y. Gim and W. Kim, “Thermodynamic phase transition in the rainbow Schwarzschild black hole,” JCAP1410, 003 (2014) [arXiv:1406.6475 [gr-qc]].
\bibitem{37}A. F. Ali, M. Faizal and B. Majumder, Europhys. Lett. 109, 20001 (2015).

\end{thebibliography}
\end{document}